\documentclass[aps,prd,eqsecnum,twocolumn,epsf,nofootinbib]{revtex4}
\usepackage{amsfonts,latexsym,amsmath,amssymb,times}
\usepackage[dvips]{graphicx}

\newcommand{\WL}{W_{t_f, }{}}

\newcommand{\tS}{\tilde{{\cal S}}}
\newcommand{\bS}{\bar{{\cal S}}}

\newcommand{\Mp}{M_{\rm pl}}
\newcommand{\bm}[1]{\hbox{\boldmath{$#1$}}}
\newcommand{\sbm}[1]{\hbox{\boldmath{\scriptsize$#1$}}}
\newcommand{\GR}[2]{G_{{\rm R}}{}^{\! #1}_{#2}}
\newcommand{\calL}[2]{{\cal L}^{#1}_{\, #2}}
\newcommand{\tr}{{\rm tr}}
\newcommand{\baralpha}{{\alpha}}
\newcommand{\balpha}{{\bar\alpha}}
\newcommand{\bbeta}{{\bar\beta}}
\newcommand{\bgamma}{{\bar\gamma}}
\newcommand{\bdelta}{{\bar\delta}}
\newcommand{\bara}{{\bar{a}}}
\newcommand{\cov}{{\cal C}}
\newcommand{\proj}{{\cal P}}
\newcommand{\Asigma}{{\tilde{\cal A}}}

\begin{document}

\thispagestyle{empty}


\title{Influence on observation from IR divergence during inflation\\
--- Multi field inflation ---}
\date{\today}
\author{Yuko Urakawa$^{1}$}
\email{yuko_at_gravity.phys.waseda.ac.jp}
\author{Takahiro Tanaka$^{2}$}
\email{tanaka_at_yukawa.kyoto-u.ac.jp}
\address{\,\\ \,\\
$^{1}$ Department of Physics, Waseda University,
Okubo 3-4-1, Shinjuku, Tokyo 169-8555, Japan\\
$^{2}$ Yukawa Institute for Theoretical Physics, Kyoto university,
  Kyoto, 606-8502, Japan}

\preprint{200*-**-**, WU-AP/***/**, hep-th/*******}


\begin{abstract}
 We propose one way to regularize the fluctuations generated during
 inflation, whose infrared (IR) corrections diverge logarithmically. 
 In the case of a single field inflation model, recently, we proposed one
 solution to the IR divergence problem. There, we introduced new
 perturbative variables which better mimic actual observable
 fluctuations, and
 proved the regularity of correlation functions with respect to these
 variables. In this paper, 
 we extend our previous discussions to a multi field inflation model. We show
 that, as long as we consider the case that the non-linear interaction
 acts for a finite duration, 
 observable fluctuations are free from IR divergences in the multi field
 model, too. In contrast to the single field model, to discuss
 observables, we need to
 take into account the effects of quantum decoherence which pick up a unique
 history of the universe from various possibilities 
 contained in initial quantum  state set 
 naturally in the early stage of the universe. 
\end{abstract}


\maketitle


\section{Introduction} \label{Introduction}
It is widely known that on the computation of the non-linear
perturbations generated during inflation we encounter the divergence
coming from the infrared (IR) corrections ~\cite{Boyanovsky:2004gq,
Boyanovsky:2004ph, Boyanovsky:2005sh, Boyanovsky:2005px, Onemli:2002hr,
Brunier:2004sb, Prokopec:2007ak, Sloth:2006az, Sloth:2006nu,
Seery:2007we, Seery:2007wf, Urakawa:2008rb}. 
As the possibility of detecting the non-linear primordial perturbations
is increasing ~\cite{Komatsu:2008hk}
\cite{Bartolo:2001cw, Bartolo:2004if, Maldacena:2002vr, Kim:2006te,
Babich:2004gb, Seery:2005wm, Seery:2005gb, Weinberg:2005vy,
Weinberg:2006ac, Rigopoulos:2005xx, Rigopoulos:2005ae, Rigopoulos:2005us, Vernizzi:2006ve,
Chen:2006nt, Battefeld:2006sz, Yokoyama:2007dw, Yokoyama:2008by,
Seery:2008ax, Naruko:2008sq, Weinberg:2008mc, Weinberg:2008nf, Weinberg:2008si, Cogollo:2008bi, Rodriguez:2008hy}, it becomes more important to solve the IR divergence
problem for the primordial perturbations and to 
predict their finite 
amplitude that we observe~\cite{Lyth:2007jh, Bartolo:2007ti, Riotto:2008mv,
Enqvist:2008kt, Urakawa:2009my}. 
In our previous work~\cite{Urakawa:2009my},
we have proposed one way to solve this IR divergence problem in the
single-field inflation model. The key observation 
is that the variables that are commonly used in describing fluctuations 
are influenced by what we cannot observe.  This is
because we can observe only the fluctuations within the region causally connected 
to us. 
We usually define the fluctuation by the deviation from the background
value which is the spatial average over the whole
universe. However, since we can observe only a finite volume of the universe, 
the fluctuations evaluated in such a way are inevitably influenced by
the information contained in the unobservable region.  
In general, the deviation from the global average is much larger than 
the deviation from the local average, which leads to the over-estimation  
of the fluctuations due to the contribution from long wavelength fluctuations.
In addition to that, to discuss the so-called observable quantities in
the framework of the standard cosmological perturbation 
(even though people call it gauge invariant formulation), in practice 
it is necessary to fix the gauge, say, the flat gauge. 
As long as the gauge is determined by solving elliptic-type equations 
on each constant time slice, 
the gauge choice is inevitably affected by the information in 
the region causally disconnected to us. 
Gauge dependence of the perturbation variables 
itself is not a problem since the ``true observables'' such
as the statistical property of the sky map of the temperature
fluctuation of the cosmic microwave background are not affected by the
gauge choice. However, if divergences appear in the quantities 
computed at the intermediate steps such as $n$-point functions of the
perturbation fields, it becomes almost impossible to extract information
about the ``true observables'' from them. 

To shut off the influence from the unobservable region of the universe,
we focused on the presence of residual gauge degrees of freedom in flat
gauge. To remove the harmful part in the residual gauge degrees of 
freedom, we imposed a further gauge condition which 
insists the local average of the inflaton to vanish. Then, the fluctuation
is not influenced by the information from the causally disconnected
region. We gave a proof that IR divergences are absent in this new scheme. In
this paper, we extend our discussion to the multi-field models. Since the
adjustment of the average value is possible only for one 
field, even if we adopt the local gauge, the fluctuations of the other
components of the scalar fields
$\varphi^I$ are still influenced by the causally disconnected
region. Here we denote the index for the D-component scalar fields
by $I = 1, \cdots, D$. Reflecting this fact, when plural fields are associated with the
IR divergences, the prescription presented in our previous paper~\cite{Urakawa:2009my} is
not enough to regularize IR divergences.    
To remove the influence from what we cannot observe, we
introduce new perturbation variables from which we can compute 
the ``true observables''. 
As the new variables, we consider the local perturbations of the fields 
defined by the deviation from their local average values. 
Then, we prove the regularity of $n$-point functions of
these new variables. 

We should note that 
$n$-point functions for the local perturbations are still influenced by what
we cannot observe. This is due to the difference between the quantum state of the
universe which we set as a natural initial condition and the one which
we observe in our real universe. The natural wave function of the
inflationary universe is not peaked at a specific point in the space of 
the local average values of $\varphi^I$.
At the observation time, this ``natural'' state of the universe can be decomposed into a 
superposition of wave packets which have a peak at a certain point. 
As the universe evolves, constituent wave packets lose correlation 
to each other. Through this so-called decoherence process,  
the coherent superposition of the wave packets starts to behave as 
a statistical ensemble of many different worlds, 
where each world means the universe described by a decohered wave 
packet~\cite{ Polarski:1995jg, Kiefer:2006je,
Starobinsky:1986fx}. 
Our observed world is just a representative one expressed by a 
wave packet randomly chosen from the various possibilities. 
Once we select one wave packet after the decoherence occurs, 
the evolution of our world will not be affected by the other 
parallel worlds.
However, the initial quantum state does include the
contributions from all the wave packets. This implies that a naive 
computation of $n$-point functions is contaminated by the other worlds uncorrelated to ours. 
In this paper, taking into account  the decoherence of the quantum state
of the universe, we propose a way to define ``observables'' 
and show that they are actually finite without suffering 
from IR divergence.

However, to be honest, the ``observables'' that we introduce 
do not correspond exactly to what we measure in the actual observations. 
Not to mislead the reader, we should stress here that our definition
of ``observables'' does not respect the aspect explained in the preceding
paragraph in a completely satisfactory manner. They are not the
expectation values for a single decohered wave packet.  
What we define is not completely free from the contamination 
of the other worlds decohered from ours. 
However, since we define our ``observables'' so as to over-estimate the 
amplitude of fluctuations, the proof of their finiteness ensures 
the finiteness of what we really observe. 
Recently the stochastic approach
~\cite{Starobinsky:1986fx, Starobinsky:1994bd} has been
employed in order to solve the IR divergence 
problem~\cite{Bartolo:2007ti, Riotto:2008mv, Enqvist:2008kt}. 
This is in harmony with our claim. 
In stochastic approach, we assume that the modes that exceed 
a certain length scale are automatically 
decohered~\cite{Starobinsky:1986fx, Starobinsky:1994bd, Nakao:1988yi,
Nambu:1988je, Morikawa:1989xz, Morikawa:1987ci, Tanaka:1997iy, Seery:2009hs}. 
Namely, long-wavelength fluctuations are treated as the statistical 
variance. Since our unique ``world'' is one realization in this
statistical ensemble, there is no contribution to the observed
quantum fluctuations from long-wavelength modes by definition.  
This is a practical way to take into account the quantum decoherence 
in the inflationary universe, but this scheme cannot completely 
remove the artificial IR cut-off scale from the discussion. 
Moreover, it is hard to deny the spiteful criticism that the reason 
why the problem of 
IR divergence does not appear in the stochastic approach might 
be simply because quantum fluctuations in the IR limit 
are neglected by hand. 
In contrast, in our approach, to avoid under-estimating the amplitude 
of fluctuations, we accepted small contamination from the 
other parallel worlds, which will make the amplitude of fluctuations
larger. 
Namely, sacrificing the accuracy of the estimate of the amplitude of fluctuations, 
we choose to show the IR regularity of the observables by
over-estimating the amplitude of fluctuations. 

This paper is organised as follows. 
In Sec.~\ref{requirement}, we give the set up of our problems. Following
it, we propose a prescription to define ``observables''. 
The basic idea of our proposal that assures the IR regularity 
is stated in this section. 
In Sec.~\ref{Proof} we explain the details of the proof of IR regularity
In Sec.~\ref{conclusion}, we summarize our results.
On the computation of the non-linear corrections, the both integrations over
the temporal and spatial coordinates can make IR corrections singular. 
In this paper, maintaining the initial time $t_i$ at a finite
past, we restrict ourselves to the time evolution of perturbations for 
a finite period of time during inflation. 
In Sec.~\ref{conclusion}, we add
the comments on the case in which we send $t_i$ to distant past. 

%
\section{A Solution to IR problem}   \label{requirement}
\subsection{Setup of the problem}
We first define the setup of the problem that we study in
this paper. We consider the multi-component inflation model with the
conventional kinetic term. The total action is given by 
\begin{eqnarray}
 S = \frac{1}{2} \int \sqrt{-g}~ [ M_{\rm pl}^2 R - {\cal G}_{IJ}
  g^{\mu\nu}\Phi^I_{,\mu}\Phi^J_{,\nu} -
  2 U(\Phi) ]d^4x~, \nonumber \\ 
\end{eqnarray}
where $M_{\rm pl}$ is the Planck mass.
We denote the field-space metric by ${\cal G}_{IJ}$. For simplicity, we
assume ${\cal G}_{IJ}$ is a constant matrix.
We perform the following change of variables
\begin{eqnarray}
 \phi_I \equiv\Phi_I/M_{\rm pl},\quad
 V(\phi)\equiv U(\Phi)/M_{\rm pl}^2, 
\end{eqnarray}
to factorize $M_{\rm pl}^2$ from the action as 
\begin{eqnarray}
 S = \frac{M_{\rm pl}^2}{2} \int \sqrt{-g}~ [R - g^{\mu\nu}\phi_{I ,\mu} \phi^I_{,\nu} 
   - 2 V(\phi) ]d^4x~,  
\end{eqnarray}
where $\phi_I\equiv {\cal G}_{IJ}\phi^I$. 
Since the Planck mass is completely factored out,
the equations of motion do not depend on it. 
The Planck mass appears only in the amplitude of quantum fluctuation. 
Namely, the typical amplitude of fluctuation of $\Phi^I$ is $H$,  
and hence that of $\phi^I$ is $H/M_{\rm pl}$. 

In order to discuss the nonlinearity, it is convenient to use the ADM
formalism~\cite{Maldacena:2002vr, Seery:2005wm, Seery:2005gb}, where the line element is expressed in terms of the lapse
function $N$, the shift vector $N^a$, and the spatial metric
$h_{ab}$:
\begin{eqnarray}
 ds^2 = - N^2 dt^2  + h_{ab} (d x^a + N^a dt) (d x^b + N^b dt)~.
\end{eqnarray}
Substituting this metric form, we can denote the action as 
\begin{eqnarray}
 S &\!\!=&\!\! 
\frac{M_{\rm pl}^2}{2} \int \sqrt{h} \Bigl[ N\,^{(3)}\!R - 2 N
  V(\phi) + \frac{1}{N} (E_{ab} E^{ab} - E^2) \cr
 && \qquad \qquad \quad  +~ \frac{1}{N} ( \dot{\phi}_I
  - N^a \partial_a \phi_I ) ( \dot{\phi}^I
  - N^a \partial_a \phi^I ) \cr 
 && \qquad \qquad \quad 
 -~ N h^{ab} \partial_a \phi_I \partial_b \phi^I \Bigr] d^4x~,
\end{eqnarray}
where
\begin{eqnarray}
 && E_{ab} = \frac{1}{2} \{ \dot{h}_{ab} - D_a N_b
 - D_b N_a \}~,  \\
 && E = h^{ab} E_{ab} ~, 
\end{eqnarray}
and $D$ is the covariant differentiation associated with $h_{ab}$.
A dot ``\,$\dot{~}$\,'' represents a differentiation with respect the time
coordinate.  
In the ADM formalism, we can obtain the constraint equations easily by varying
the action with respect to $N$ and $N^a$, 
which play the role of
Lagrange multipliers. 
We obtain the Hamiltonian constraint
equation and the momentum constraint equations as  
\begin{eqnarray}
 && ^{(3)} R - 2 V -  N^{-2} (E^{ab} E_{ab} - E^2 ) \cr
 && \qquad
 -~  N^{-2} ( \dot{\phi}^I - N^a \partial_a \phi^I)
 ( \dot{\phi}_I - N^a \partial_a \phi_I) \cr
 && \qquad
 -~ h^{ab}
 \partial_a \phi^I \partial_b \phi_I = 0~,   \\
 && D_a [ N^{-1} ( E^a_{~b} - \delta^a_{~b} E )] - N^{-1} 
  \partial_b \phi_I ( \dot{\phi}^I - N^a \partial_a \phi^I) = 0~. 
 \nonumber \\
\end{eqnarray}
Hereafter, neglecting the vector perturbation, we denote the shift vector as
$N_a = \partial_a \chi$. 
In this paper we work in the flat gauge, defined by 
\begin{eqnarray}
 h_{ab} = e^{2 \rho }  \delta_{ab}~, \label{GC flat}
\end{eqnarray}
where $a\equiv e^{\rho}$ is the background scale factor.  
Here we have also neglected the tensor perturbation, 
focusing only on the scalar perturbation, in which 
the IR divergence of our interest arises~\cite{Boyanovsky:2005px,
Urakawa:2008rb}.

In this gauge, using $N$, $\chi$ and the fluctuation of the
scalar fields $\varphi^I$, the total action is written as 
\begin{eqnarray}
 S &\!=&\! \frac{M_{\rm pl}^2}{2} \int  dt\, d^3\! {\bm x}\, e^{3\rho}
  \Bigl[  - 2 N~ 
 \displaystyle \sum_{n=0} \frac{1}{n!} V_{I_1 \cdots I_n}(\phi)
   \varphi^{I_1} \cdots \varphi^{I_n} \nonumber \\ &&~
  + N^{-1} \{ - 6 \dot{\rho}^2 + 4 \dot{\rho}~ 
   \triangle \chi  \nonumber \\ && \qquad \qquad \qquad  \qquad
 + ( \nabla^a \nabla^b \chi \nabla_a
   \nabla_b \chi - (\triangle \chi)^2 ) \} \nonumber \\ && ~
  + N^{-1}  ( \dot{\phi}^I + \dot{\varphi}^I - \nabla^a \chi \nabla_a
  \varphi^I) ( \dot{\phi}_I + \dot{\varphi}_I - \nabla^a \chi \nabla_a
  \varphi_I)  \nonumber \\ && \qquad \qquad \qquad  \qquad
  - N \nabla_a \varphi_I \nabla^a \varphi^I \Bigr]~, 
 \label{action in flat}
\end{eqnarray}
and two constraint equations are 
\begin{eqnarray}
 && 2  N^2 \displaystyle \sum_{n=0} \frac{1}{n!} 
  V_{I_1 \cdots I_n}(\phi)
   \varphi^{I_1} \cdots \varphi^{I_n} - 6 \dot{\rho}^2 \cr && \qquad
 +~ 4 \dot{\rho} \triangle \chi 
  + \{ \nabla^a \nabla^b \chi \nabla_a
   \nabla_b \chi - (\triangle \chi)^2 \} 
 \cr  &&  \qquad
   +~ ( \dot{\phi}_I + \dot{\varphi}_I - \nabla^a \chi
   \nabla_a \varphi_I) ( \dot{\phi}^I + \dot{\varphi}^I - \nabla^b \chi
   \nabla_b \varphi^I) \cr  &&  \qquad
   +~ N^2 \nabla_a \varphi_I \nabla^a \varphi^I   = 0,  \nonumber \\ \\
&& (\nabla_a  N) \{ 2 \dot{\rho} \delta^a_{~b} +
   ( \nabla^a \nabla_b \chi - \delta^a_{~b} \triangle
  \chi ) \} \cr && \qquad
 -  (\nabla_b \varphi_I) N ~ ( \dot{\phi}^I + \dot{\varphi}^I - 
    \nabla^a \chi \nabla_a \varphi^I) = 0 ~, \nonumber \\
\end{eqnarray}
where 
$$
\nabla_a\equiv e^{-\rho}\partial_a~,
$$ 
represents the three dimensional partial
differentiation with respect to the proper length 
coordinates ${\bm X} \equiv e^{\rho}{\bm x}$ and
$$
\triangle \equiv \delta^{ab}\nabla_a\nabla_b~.
$$ 
Spatial indices, $a, b, \cdots$, are 
raised by $\delta^{ab}$. 
We use this notation, which respects the proper distance, 
because it eliminates 
all the complicated scale factor dependences from the action. 
We define the derivatives of the potential as
\begin{eqnarray}
 V_{I_1  I_2 \cdots I_n}(\phi) \equiv \frac{\partial^n V(\phi)}{\partial
  \phi^{I_1} \partial \phi^{I_2} \cdots \partial  \phi^{I_n} }~.
\end{eqnarray}
The background quantities $\rho$ and $\phi$ satisfy the following equations :
\begin{eqnarray}
 && 3 \dot{\rho}^2  = \frac{1}{2}  \dot{\phi}^I \dot{\phi}_I + V(\phi)~, \\
 && \ddot{\phi}^I + 3 \dot{\rho} \dot{\phi}^I +  V^I = 0~, \\
 && \ddot{\rho} = - \frac{1}{2}  \dot{\phi}^I \dot{\phi}_I~. 
\end{eqnarray}
Expanding the variables as
\begin{eqnarray*}
 N &=& 1 + \delta N_1 + \frac{1}{2} \delta N_2 + \cdots~, \cr
 \chi &=& \chi_1 + \frac{1}{2} \chi_2 + \cdots~, \nonumber\\
 \varphi^I &=& \varphi^I_1 + \frac{1}{2} \varphi^I_2 + \cdots~,
\end{eqnarray*}
we find that the first order constraint equations are
\begin{eqnarray}
 && V_I \varphi_1^I + 2V \delta N_1 + 2 \dot{\rho} \triangle^2 \chi_1~,
 + \dot{\phi}_I \dot{\varphi}^I_1 = 0~,
 \label{Hconst 1} \\
  && \nabla_a  \left( 2 \dot{\rho}~ \delta N_1 -
  \dot{\phi}_I \varphi^I_1 \right) = 0~.   \label{Mconst 1}
\end{eqnarray}
The first order perturbation $\varphi^I_1$ is identified with 
the field perturbation in the interaction picture. 
Taking the deviation of the action with respect to $\varphi_I$, we can
derive the equation of motion for $\varphi_I$, which includes the Lagrange
multipliers $\delta N$ and $\chi$. For example, from the third
order action, we can derive the equation of motion with quadratic
interaction terms as follows,
\begin{widetext}
\begin{eqnarray}
 && \ddot{\varphi}^I + 3 \dot{\rho} \dot{\varphi}^I - \triangle
  \varphi^I + V^I_{\, J} \varphi^J -   \dot{\phi}^I \triangle
 \chi + \delta N V^I - 3 \dot{\rho}~ \dot{\phi}^I \delta N - \partial_t (
 \delta N \dot{\phi}^I) \nonumber \\
 && \qquad
   +~ \frac{1}{2} V^I_{\, JK} \varphi^J \varphi^K
   - \nabla_a ( \dot{\varphi}^I - \dot{\phi}^I \delta N )
   \nabla^a \chi -   ( \dot{\varphi}^I - \dot{\phi}^I \delta N )
  \triangle^2 \chi  
   - \dot{\rho} \nabla^a \chi \nabla_a
  \varphi^I -  \partial_t ( \nabla^a \chi \nabla_a \varphi^I
  )  \nonumber \\ && \qquad
 -~  3 \dot{\rho}~ \dot{\varphi}^I \delta N - \partial_t (\delta N
 \dot{\varphi}^I)
 - \nabla_a( \delta N \nabla^a \varphi^I )
 + V^I_{\, J} \varphi^J \delta N + 3 \dot{\rho} \dot{\phi}^I \delta N^2 
 + \partial_t ( \dot{\phi}^I \delta N^2 ) = 0
 \nonumber \\  \label{eom for varphi}
\end{eqnarray}
\end{widetext}
Solving the constraint equations at each order, we can express the
lapse function and the shift vector as functions of $\varphi^I$ at lower
order. Substituting
these expressions into the equation of motion for $\varphi^I$, which is
up to third order given by Eq.~(\ref{eom for varphi}), the equation is written solely in terms of the
dynamical degree of freedom, $\varphi^I$. 

\subsection{Tree-shaped graphs}  \label{tree diagram}
In this subsection, as a preparation for computing $n$-point functions of $\varphi_I(x)$, we
consider an expansion of the Heisenberg field $\varphi_I(x)$ in terms of the interaction
picture field, 
using the retarded Green function $\GR{I}{J}(x,x')$, 
which is causal. 
Since the retarded Green function $\GR{I}{J}(x,x')$
has a finite non-vanishing support for fixed $x$ and $t'$,  
its three dimensional Fourier transform with respect to $\bm{x}'$
becomes regular in 
the IR limit.

Let us denote the equation of motion for $\varphi^I$ schematically by 
\begin{eqnarray}
{\cal L}^I_{\,J} \varphi^J =- \Gamma^I [\varphi]~, 
 \label{schematical eom}
\end{eqnarray}
where $\calL{I}{J}$ 
is a second order differential operator 
corresponding to the linearized equation for $\varphi^J$ 
(Eq.~(\ref{Eq:starteq})) and 
$\Gamma^I$ stands for all the nonlinear interaction terms. 
Using the retarded Green function $\GR{J}{K} (x, x')$ that satisfies   
\begin{eqnarray}
\calL{I}{J}
\GR{J}{K}(x, x') = -a^{- 3} \delta^4(x - x') \delta^I_{\,K} ~,
 \label{eom GR}
\end{eqnarray} 
we can solve Eq.~(\ref{schematical eom}) formally as 
\begin{eqnarray}
 \!\! \varphi^I(x) = \varphi^I_1(x) + \int d^4\!x'\,\GR{I}{J}(x, x') a^{3}(t')
  \Gamma^J[\varphi](x'). 
 \label{general solution varphi}
\end{eqnarray}
where 
the first order perturbation $\varphi^I_1$ satisfies
\begin{eqnarray}
 \calL{I}{J} \varphi^J_1 (x) = 0~. 
\end{eqnarray}
Here the factor $a^{3}$ originates from the
background value of $\sqrt{-g}$.
Substituting the expression (\ref{general solution varphi}) for $\varphi^I(x)$ 
iteratively into $\Gamma_I[\varphi]$  on its r.h.s., 
we obtain the Heisenberg field $\varphi^I(x)$ 
expanded in terms of $\varphi^I_1(x)$ to an arbitrary high order 
using the retarded Green function $\GR{I}{J}(x, x')$. 
As we have shown in \cite{Urakawa:2009my}, to expand $\varphi^I$,
a diagrammatic illustration will be useful. 
The Heisenberg field can be
expressed by a summation of tree-shaped graphs in which
all the retarded Green functions $\GR{I}{J}(x, x')$ are followed
by two or more $\varphi^I_1(x')$ or $\GR{I}{J}(x', x'')$ with some
integro-differential operators and all the interaction picture fields
$\varphi^I_1(x)$ are located at the right most ends of the graphs.

When we compute the expectation value for $n$-point functions  
of the Heisenberg field, the interaction picture fields $\varphi_I$
in the tree-shaped graphs are 
contracted with each other to make pairs, which 
are replaced with Wightman functions, 
$G_+^{IJ}(x, x')\equiv \langle \varphi^I_1(x)\varphi^J_1(x') \rangle$ 
or $G_-^{IJ}(x, x')(=G_+^{JI}(x', x))$. 
These propagators are IR singular ($\propto 1/k^3$), 
which is the possible origin of IR divergences in momentum integrations.  
While, the retarded Green function 
\begin{eqnarray}
 \GR{I}{J}(x, x') = i \theta(t - t') M^2_{\rm pl} \{G^{~~I}_{+\, J}(x, x') -
  G^{~~I}_{-\,J}(x, x') \}~,\cr 
 \label{Def:GR}
\end{eqnarray}
is regular in the IR limit.

\subsection{Iteration scheme and local gauge conditions}  \label{guage fixing}
In our previous paper \cite{Urakawa:2009my}, we have shown that the
flat gauge still has residual gauge degrees of freedom. 
For instance, we can introduce
an arbitrary function $f_n(t)$ to the $n$-th order lapse function and the
shift vector as
\begin{eqnarray*}
 \delta N_n ~\rightarrow ~ \delta N_n + f_n(t), \quad
 \chi_n ~ \rightarrow ~ \chi_n - \frac{V}{6 \dot{\rho}} X_a X^a f_n(t).
\end{eqnarray*}
This gauge degree of freedom
corresponds to the scale transformation. As is mentioned in \S~\ref{Introduction}, our final goal is to 
define finite observable quantities in place of the naively 
divergent quantum correlation functions. 
For this purpose, 
we need to define gauge invariant 
variables without the information 
contained in the region far outside ${\cal O}$. 
Then, we have to fix the residual gauge only using the information 
near the observable region ${\cal O}$.

In the multi-field model, it is convenient to decompose the
perturbation into the adiabatic one, 
which is tangential to the background trajectory,
and the entropy one, which is orthogonal to the background
trajectory  \cite{Gordon:2000hv}.  Using the residual gauge degrees of freedom, we fix the
homogeneous mode in the direction of the background trajectory
$e_I \equiv \dot{\phi}_I / (\dot{\phi}_J \dot{\phi}^J)^{\frac{1}{2}}$
as
\begin{eqnarray}
  \hat W_t~ e_I \tilde{\varphi}^I (t) \equiv \frac{1}{L_t^3}
   \int d^3 \bm{x}~W_t (\bm{x})~ e_I
   \tilde{\varphi}^I(t,~\bm{x})  = 0,  \label{local gauge}
\end{eqnarray}
where $W_t (\bm{x})$ 
is a window function, which 
is unity in the finite region 
${\cal O}_t \equiv {\cal O} \cap \Sigma_t$ 
with a rapidly vanishing halo in the surrounding region, 
where $\Sigma_{t}$ means a $t=$const. hypersurface corresponding to the
time $t$.
For definiteness, we introduce 
${\cal O}'_{t_f}\supset {\cal O}_{t_f}$ and 
define ${\cal O}'$ as the causal past of ${\cal O}'_{t_f}$. 
We require $W_t (\bm{x})$ to vanish in the region outside ${\cal O}'$. 
In addition, $W_t (\bm{x})$ is supposed to be a sufficiently smooth function 
so that an artificial UV contribution is not induced by a sharp cutoff. 
$L_t$, an approximate radius of the
region ${\cal O}_t$, is defined such that the normalization condition 
$$
\hat W_t 1= 1~,
$$ is satisfied.

We associated ``$\,~\tilde{}~\,$'' 
with the variables in the particular gauge satisfying Eq.~(\ref{local gauge}), 
in order to clearly distinguish them from the variables 
for which the additional gauge condition
is not imposed. 
The difference between the variables with and without ``$\,~\tilde{}~\,$''  is 
only in the boundary conditions. Hence, they 
obey the same differential
equations, (\ref{Mconst 1})-(\ref{eom for varphi}).

In order to fix the arbitrary functions $f_n(t)$
($n=1, 2, 3, 4, \cdots$) 
so as to satisfy the gauge condition~(\ref{local gauge}), 
we need to obtain a formal solution for
$\tilde\varphi$.
The higher order lapse functions are determined by solving 
the momentum constraint given in the form 
\begin{eqnarray}
\nabla_a\left(\delta \tilde N_n-{1\over
	   2\dot\rho} \dot\phi_I \varphi^I_n\right)
      =\Xi^{(n)}_{a},\quad (n=1,2,3,\cdots),~
\label{lapseEq}
\end{eqnarray}
where the r.h.s.~is the $n$-th order 
nonlinear term expressed in terms of the lower order lapse functions, shift
vectors, and $\tilde\varphi$.  
As we neglect the vector perturbation, 
we consider only the scalar part of these equations, i.e.
its divergence, which is formally solved as
\begin{eqnarray}
\delta \tilde N_n= \delta \breve N_n+f_n, 
\label{Eq:Ni}
\end{eqnarray}
with
\begin{eqnarray*} 
\delta \breve N_n={1 \over 2\dot\rho} \dot\phi_I \varphi^I_n
+\triangle^{-1} \nabla^a\Xi^{(n)}_{a}.
\label{Eq:Nn}
\end{eqnarray*}
We define the operation $\triangle^{-1}$ by
\begin{eqnarray}
 \triangle^{-1} F(x) = -  \frac{1}{4 \pi} \int
  \frac{W_t(e^{-\rho}{\bm Y}) d^3 {\bm Y}}{|{\bm X} - {\bm Y}|} F(t,e^{-\rho}{\bm Y})~, 
 \label{def DF}
\end{eqnarray}
so that it is completely determined by the local 
information in the neighborhood of ${\cal O}_t$. 
Similarly, the higher order shift vectors satisfy the Hamiltonian constraint 
in the form 
\begin{eqnarray*}
\triangle \tilde \chi_n
\!&=\!& {1\over 2}\left[
       {\ddot\phi_I}\,\varphi^I_n - 
       {\dot\phi_I\over\dot\rho^2}
       \partial_t \left( {\dot\rho \varphi^I_n }\right)
\right]
 \nonumber \\ && \qquad \qquad  -{V\over
 \dot\rho}\left(f_n+\triangle^{-1} \nabla^a\Xi^{(n)}_{a} \right) + C_n,
\end{eqnarray*}
where $C_n$ on the r.h.s.~is a function expressed in terms 
of the lower order lapse functions, shift vector and $\tilde\varphi^I$. 
A formal solution for $\tilde \chi_n$ is given by 
\begin{eqnarray}
\tilde \chi_n&\!\!=&\!\! 
 \breve\chi_n -{r^2 V\over 6\dot\rho} f_n~,
\label{Eq:chii}
\end{eqnarray}
with
\begin{eqnarray*}
\breve\chi_n &=& 
\triangle^{-1} 
  \Biggl({1\over 2}\left[
       {\ddot\phi_I}\,\varphi^I_n - 
       {\dot\phi_I \over\dot\rho^2}
       \partial_t \left({\dot\rho \varphi^I_n}\right)\right]
 \nonumber \\ && \qquad \qquad \qquad \qquad
      - {V\over \dot\rho}\triangle^{-1}
       \left(\nabla^a\Xi^{(n)}_{a}\right)+C_n \Biggr)~.
\label{Eq:chin}
\end{eqnarray*}
Substituting the expressions for the lapse function (\ref{Eq:Ni})
and the shift 
vector (\ref{Eq:chii}) into the equation of motion for $\tilde\varphi$ 
truncated at the $n$-th order, we obtain an equation 
\begin{equation}
\calL{I}{J}~ \tilde\varphi^J_n
-\dot\phi^I \dot f_n+ \left({V\dot\phi^I\over\dot \rho}+2V^I \right)f_n
=-W_t({\bm x})\Gamma^I_n,  
\label{eomformal}
\end{equation}
where, for later convenience, 
we have inserted a window function $W_t({\bm x})$ on the r.h.s., which 
does not alter the evolution in ${\cal O}$.  
The explicit form of $\calL{I}{J}$ is given by 
\begin{equation}
\calL{I}{J} \equiv (\partial^2_t +  3 \dot\rho\, \partial_t -
 \triangle) \delta^I_{\,J}
 +\left(V^I_{\, J} - e^{-3 \rho} \dot A^I_J\right).
\label{Eq:starteq}
\end{equation}
with
$$
A^I_{J}(t)\equiv e^{3\rho}\dot\phi^I \dot\phi_J/\dot\rho~, 
$$
and $\Gamma^I_n$ on the r.h.s.~of Eq.~(\ref{eomformal}) 
represents all the $n$-th order nonlinear terms 
expressed in terms of lower order variables. 

The equation for the inhomogeneous part of 
$\tilde\varphi_n^I$ is obtained 
by acting the operator 
$1-\hat W_t$ on Eq.~(\ref{eomformal}) 
as 
\begin{eqnarray}
(1-\hat W_t){\cal L}_{IJ} \tilde{\varphi}_n^J = - 
(1-\hat W_t) W_t({\bm x})\, 
 \Gamma_{I,n}[\tilde{\varphi}]~.
 \label{Eq:varphiad}
\end{eqnarray}
Then, we find that $\tilde\varphi^I_n(x)$ is obtained by
\begin{eqnarray}
\tilde\varphi^I_n(x)\equiv 
 \hat{\bar W}^{\, I}_{t\, J} 
 \breve\varphi^J_n(x)+B_{\perp n}^{I}(t)~, \label{Sol:Tvarphin}
\end{eqnarray}
where $\hat{\bar W}^{\, I}_{t\, J} 
\equiv \delta^I_J - e^I e_J \hat{W}_t$  and
$\breve\varphi^I_n$ satisfies 
\begin{eqnarray}
\calL{I}{J} \breve\varphi^J_n(x) = -W_{t}({\bm x})
\Gamma^I_n[\tilde{\varphi}]~.
\label{breve}
\end{eqnarray}
$B_{\perp n}^I(t)$ 
is a homogeneous field perpendicular to $e^I$. 
The solution~(\ref{Sol:Tvarphin}) satisfies the 
gauge condition $\hat W_t\, e_I \tilde\varphi^I_n=0$. 

The remaining unknowns are $f_n(t)$ and $B_{\perp n}^I(t)$ which has 
$D-1$ components. These $D$ unknown components are 
determined by the homogeneous part of 
the equations of motion obtained by substituting 
(\ref{Sol:Tvarphin}) into Eq.~(\ref{eomformal}), 
\begin{equation}
\calL{I}{J}B^J_{\perp n}
-\dot\phi^I \dot f_n+ \left({V\dot\phi^I\over\dot \rho}+2V^I \right)f_n
=\calL{I}{J}e^J e_K\hat W_t\breve\varphi^K_n.  
\label{Bfeq}
\end{equation}

\subsection{Projection to one decohered wave packet} \label{projection}
When plural fields have scale invariant or even redder
spectrum, the entropy perturbation can give divergences.
However, in this subsection we show that a naive computation of 
the correlation functions does not give the
correlation functions that we actually observe. 

When there is no isocurvature mode related to IR divergence, 
making use of the gauge degree of freedom, 
we can arrange that the adiabatic perturbation variable 
$\Asigma\equiv e_I\tilde\varphi^I$ 
to be the deviation from the local average value. 
In contrast, there is no such gauge degree of freedom 
for the isocurvature perturbation 
$$
{\cal S}^I\equiv \tilde\varphi^I-e^Ie_J\tilde\varphi^J.
$$
Hence, we have to use the isocurvature perturbation variables 
defined by the deviation from the average values on a whole time
slice, which contains information of the causally disconnected region.  
As observable isocurvature perturbation,
we introduce the local fields,
\begin{eqnarray}
 \tS^I(x) \equiv {\cal S}^I (x) - \hat{W}_t {\cal S}^I(x)~.
\label{def varphiobs}
\end{eqnarray}
However, even if we restrict our attention to the local quantity
$\tS_I(x)$ on the final surface, 
the variables ${\cal S}_I(x)$ which contains the information outside the 
causal region appear in describing time evolution of the field. 
Although in our previous work we have stressed that the locality of the
observables is the key issue in order to assure the IR regularity, 
the locality is inevitably violated under the 
presence of IR divergence originating from isocurvature perturbation.

Here we need to raise another key issue, i.e. the quantum decoherence. 
The primordial perturbations are expected to decohere through 
the cosmic expansion and/or through various
interactions~\cite{Polarski:1995jg, Kiefer:2006je, Starobinsky:1986fx}. 
This decoherence process transmutes the quantum fluctuations 
at a long wavelength to statistical variances~\cite{Starobinsky:1986fx,
Starobinsky:1994bd, Nakao:1988yi, Nambu:1988je, Morikawa:1989xz,
Morikawa:1987ci, Tanaka:1997iy, Roura:2007jj, Urakawa:2007dm}.  
At the initial time when the wavelength of relevant modes is short, 
the adiabatic vacuum state will be a natural vacuum state.  
However, the adiabatic vacuum state is not a wave packet 
sharply peaked around a specific value of the homogeneous part of 
the scalar field $\hat{W}_t \tilde\varphi^I$. 
Instead, it is infinitely broad and can be interpreted as 
a coherent superposition of peaked wave packets. 
(Detailed explanation will be given in \S~\ref{Bogoliubov} below.) 
In the early stage of inflation these 
wave-packets correlate to each other, but the quantum coherence is 
gradually lost in the course of time evolution. 
Thus, at the observation time ($t=t_f$) the coherence will 
remain only between adjacent overlapping wave packets. 
Our observed world is corresponding to one decohered wave packet 
picked up from this superposition. 
For the later time evolution of our world, we can completely neglect 
the other wave packets whose peak is located very far 
from ours in the space of isocurvature components of the 
local average values of fields $\tS^I\equiv
(\hat{W}_{t_f}\tilde\varphi^I)_{\perp}$. 
Hence, keeping the superposition of all wave packets 
as the wave function of the universe 
gives rather misleading results, i.e. huge over-estimates 
of quantum fluctuations. 
We should therefore remove the contributions from the other worlds. 


It is standard to discuss the decoherence process by coarse-graining some
degrees of freedom in the quantum interacting system, by which 
the reduced density matrix evolves from its initial pure state to 
a mixed state. 
This process is interpreted as the transition from the 
initial coherent superposition of many different worlds to the final 
statistical ensemble of them. 
In \cite{Bartolo:2007ti, Riotto:2008mv, Enqvist:2008kt},
the decoherence process of the long wavelength modes is treated 
by using the stochastic approach to inflation, in which 
all the quantum fluctuations of long wavelength modes 
are assumed to turn into the variance of classical ensemble at each 
time step. 
This assumption of complete classicalization can be justified 
to some extent by coarse-graining the short wavelength modes. 
On physical ground, we believe that this approach gives a good 
approximate description of the dynamics of inflation. \cite{Finelli:2008zg}
However, here we take a different approach because  
in the stochastic approach, by assumption, the quantum fluctuations 
of long wavelength modes, which we focus on in the present paper,  
cannot enter into the quantum loop corrections 
from the beginning. 
In this regard we think that stochastic approach is not much more 
satisfactory compared with a naive prescription of introducing a cutoff 
length scale by hand.

%
The accurate evaluation of what we really observe requires to
elucidate the decoherence process of the primordial perturbations, 
which is a long-lasting and unsettled issue. 
Here, instead, we aim at proving the IR regularity of our ``observables'' 
independently of the details of the decoherence process, which is the
heart of this paper. 
Although it is difficult to understand how the 
decoherence process proceeds until the observation time
$t_f$, it is natural to expect that 
the wave function of the universe has been already decohered at $t = t_f$ 
to a large extent.  
The observation picks up one world from the superposition of many 
decohered worlds.
The wave function corresponding to each decohered world 
will have a rather sharp peak in the coordinate space of
\begin{eqnarray}
 \bS^\balpha (t)\equiv e^{\balpha}_I \hat{W}_t {\cal S}^I(x),~\quad (\balpha=2,3,\cdots,D),
\end{eqnarray} 
where $\{e^{\alpha}_I\}$ with $e^{1}_I=e_I$ 
is a set of orthonormal bases in field space.  
Hence, we insert a projection operator $\proj_{\{\alpha\}}$ which restricts the values of
$\bS^\balpha (t_f)$ 
to a small range near $\bS^\balpha (t_f)=\alpha^\balpha$ 
without making any significant effect on 
each decohered wave packet.
Then, the insertion of $\proj$ is expected to reduce 
the contamination from the other worlds significantly. 

For simplicity, we choose the Gaussian projection
operator
\begin{eqnarray}
 \proj_{\{\alpha\}} 
\equiv \exp \Bigl[ -
{\cov^{-1}_{\alpha\beta}(\bS^\balpha(t_f)-\alpha^\balpha)
   (\bS^\bbeta(t_f)-\alpha^\bbeta)\over 2}\Bigr]~,\cr
\label{gaussianproj}
\end{eqnarray}
where $\alpha^\balpha$ are $D-1$ real C-numbers
\footnote{Here we used the word ``projection operator'', but 
this operator does not satisfy the relation 
$\proj_{\{\alpha\}}= \proj^2_{\{\alpha\}}$ expected from its name. 
However, this kind of property is unnecessary for our present discussion.}.
The dispersion should be sufficiently large compared with 
the width of one decohered wave packet to guarantee that the evaluated 
amplitude of fluctuations is always larger than that for 
a single wave packet. 
Inserting an identity 
\begin{eqnarray}
 {1\over \sqrt{\det\cov}}
 \left[ \prod_{\balpha=2}^D \int^{\infty}_{- \infty} \frac{d \alpha^{\balpha}}{\sqrt{2\pi}} 
   \right]
\proj_{\{\alpha\}}
  =  1~,  \label{identity P}
\end{eqnarray}
we can expand the $n$-point function of 
the variables whose local average values are subtracted. 
$\{\Asigma, \tS^I\}$ 
are schematically denoted by $\tilde{O}$. 
We can expand the $n$-point function of 
$\tilde{O}(x)$ for the adiabatic vacuum $~|~0~ \rangle_a$ as 
\begin{eqnarray}
&&\!\!\!\!\!\!\!\!
   {}_a \langle 0| \tilde{O} (t_f, {\bm x}_1)
  \tilde{O}(t_f, {\bm x}_2) \cdots \tilde{O}(t_f, {\bm x}_n) |0 \rangle_a 
  \nonumber \\ &&\!\! 
 = {1\over \sqrt{\det\cov}}
 \left[ \prod_{\balpha=2}^D
   \int^{\infty}_{- \infty} \frac{d \alpha^{\balpha}}{\sqrt{2\pi}} 
   \right]\cr
&&\quad\times
  {}_a \langle 0|\proj_{\{\alpha\}} \tilde{O}(t_f, {\bm x}_1)  \cdots
  \tilde{O}(t_f, {\bm x}_n) |0 \rangle_a~.
\label{corre with P0}
\end{eqnarray}
Then, we regard  
\begin{eqnarray}
&&  \langle\, \proj\, \tilde{O}(t_f, {\bm x}_1) \tilde{O}(t_f, {\bm x}_2)
 \cdots \tilde{O}(t_f, {\bm x}_n) \, \rangle \nonumber \\
&&~~ \equiv 
{ _a \langle\, 0\,|\proj \tilde{O}(t_f, {\bm x}_1) \tilde{O}(t_f, {\bm x}_2)
 \cdots \tilde{O}(t_f, {\bm x}_n) |\,0\, \rangle_a  
\over 
 _a \langle\, 0\,|\proj |\,0\, \rangle_a
}, \nonumber \\
\label{obs npointfn}
\end{eqnarray}
in the integrand on the right hand side of Eq.~(\ref{corre with P0}) 
as the observable $n$-point function of $\tilde{O}$s after the 
selection of a single decohered world. 
Here we set $\alpha^\balpha=0$ 
and denote $\proj_{\{\alpha^\balpha=0\}}$ by $\proj$. 
Setting $\alpha^\balpha=0$ does not lose generality because the 
classical average values of isocurvature perturbation $\bS^\balpha$ can be changed by
choosing the background trajectory. 
We will prove the IR regularity of this $n$-point function
in the succeeding section. 

Now the question is how to determine the width of the projection
operator, $\sigma$. (Here we are assuming that 
${\cal C}^{\balpha\bbeta}\approx \sigma^2 \delta^{\balpha\bbeta}$.) 
On one hand, $\sigma$ must be chosen large enough to exceed the 
width of a decohered wave packet. 
Naively there is a minimum size of the wave packet since 
a very narrow wave packet cannot maintain its width for a long period of 
time. 
Later, we find that the minimum size of a wave packet that we can choose 
is determined by the typical amplitude of quantum fluctuations
generated during inflation, which is characterized by the Hubble scale
for a nearly massless scalar field. 
This amplitude is $H/\Mp$ in terms of the fluctuation of $\tilde{O}(x)$. 
Therefore we need to set $\sigma$ to be larger than $H/\Mp$. 
On the other hand, in order to suppress the
influence from other wave packets, $\sigma$ should
not be very large. Later, we find that the condition  
that the higher order contributions are more suppressed 
requires $\sigma$ to be much less than unity. 
These two conditions are compatible by choosing $\sigma$ 
to satisfy $H /\Mp \ll \sigma \ll 1$.

Due to the inaccurate evaluation of the decohered wave packet, 
the effect of insertion of the Gaussian projection 
is not equivalent to selecting our world through 
the actual decoherence process. Hence, we cannot claim that the
$n$-point function given by Eq.~(\ref{obs npointfn}) is the true
observable $n$-point function. 
However, the former amplitude is larger than the latter one. 
Thus, if the $n$-point function given by Eq.~(\ref{obs
npointfn}) is proved to be finite, we can conclude 
that the $n$-point function of $\tilde{O}(x)$ evaluated for 
the actual decohered wave packet is also finite.

In the above discussion we assumed that the average values $\bS^\balpha(t)$ of all
entropy modes has accomplished decoherence process successfully
before we measure $n$-point functions of $\tilde O(x)$ at $t=t_f$.
Here we want to stress that
whether the superposition of decohered wave packets come to be  
statistical ensemble
or not has nothing to do with whether the mode is measurable for us or  
not.
Let's consider a hidden variable $x$ which interacts extremely
weakly with our visible sector. Even in that case, if $x$ represents
an average of a field over a large volume,
the quantum coherence between two wave packets $|1\rangle$ and $|2\rangle$
peaked at largely different values of $x$
will be lost (at least after integrating out the other degrees of freedom
in the hidden sector). Assuming that $x$ takes the two discrete state
$|1\rangle$ and $|2\rangle$ with an equal weight for simplicity,
the evolved density matrix will be schematically written as
$\rho=(|1\rangle\langle 1|\rho_1+|2\rangle\langle 2|\rho_2)/2$,
after integrating out the other degrees of freedom in the hidden sector.
Here $\rho_1$ and $\rho_2$ are the density matrices of our visible sector.
(If the interaction between the hidden and visible sectors
is extremely weak, $\rho_1$ and $\rho_2$ are identical.)
Then, for any operator ${\cal O}$ in our visible sector,
$\tr \rho {\cal O}=(\tr\rho_1{\cal O}+\tr\rho_2{\cal O})/2$.
This means that, as long as the variables measurable for us are
concerned, the state can be understood as a statistical
ensemble composed of $\rho_1$ and $\rho_2$. Therefore what we actually
observe is the expectation value for either $\rho_1$ or $\rho_2$.
Therefore, irrespective of whether the isocurvature perturbation
is in the visible sector or in the hidden sector,
we can insert a projection operator $\proj$
to take into account the influence of decoherence.
In the succeeding section, we discuss
the regularity of the ``observed'' $n$-point function
$\langle \proj\, \tilde{O} (t_f, {\bm x}_1)
 \tilde{O}(t_f, {\bm x}_2) \cdots \tilde{O}(t_f, {\bm x}_n)  \rangle$.

\section{Proof of IR regularity}  \label{Proof}
In this section, we prove the IR regularity of the $n$-point function
Eq.~(\ref{obs npointfn}). In this paper, we discuss the evolution of 
perturbation during a finite period of inflation. 
First we describe the way of the quantization in
 \S\ref{Quantization}.  Before
 starting the detailed discussion, in \S\ref{Sketch}, we briefly 
explain the basic idea of the proof of
 IR regularity. In this subsection, we clarify the difference between the
 regularization in multi-field models and that in single field
 models. After that, in \S \ref{Squeezed}, we adapt the basis
 transformation. In the new basis, it becomes easier to understand the
 regularization in the multi-field models.  
Based on these preparations, in \S
 \ref{Projection}, we show that IR suppression due to the projection operator
 regularizes the IR divergence when the initial conditions are set 
at a finite past.

In this section, we discuss the IR regularity after we remove
the influence of the unobservable quantities.
For the technical reason, it is better to avoid treating the divergent
quantities directly. Therefore, first, we assume that the total volume 
of the universe $V_c = L_c^3$ is finite.  
Then, the normal modes take discrete spectrum, and as a result 
IR divergence is concentrated on the spatially homogeneous 
mode with $p=0$, as long as $L_c$ is kept finite. 
Even with a finite volume, 
the quantum fluctuation of the homogeneous mode with $p=0$ in 
adiabatic vacuum is still divergent in contrast to the other IR modes. 
In Sec.~\ref{Quantization} we introduce a parameter $s_{\baralpha}$ that measures 
the deviation from the adiabatic one for the $p=0$ mode. 
At the end of calculations, we take the limit 
$V_c\to\infty$ and $s_{\baralpha} \to 0$.  

\subsection{Quantization}  \label{Quantization}
In the previous section, we described how we can expand the
Heisenberg field $\tilde{\varphi}^I$ in terms of
$\breve{\varphi}^I_1(x)$. The interaction picture field 
$\breve{\varphi}^I_1(x)$
\footnote{The leading term of the Heisenberg picture field 
$\breve{\varphi}^I_1(x)$ agrees with the interaction picture
field $\breve{\varphi}^I_{\rm int}(x)$. Thus, we denote the interaction picture
field as $\breve{\varphi}^I_1(x)$.}
satisfies the equation of motion
${\cal L}_{IJ} \breve \varphi^J_1 (x) = 0$.
Using a set of mode function
$ \{ \phi^I_{~\alpha, \sbm{p}} (x) \equiv u^I_{~\alpha,p}(t) e^{i {\sbm
p} \cdot {\sbm x}} \}$ which satisfies
\begin{eqnarray}
 0 &=& e^{- i {\sbm p} \cdot {\sbm x}} \calL{I}{J} 
   \phi^J_{~\alpha,{\sbm p}}
  \nonumber \\
   &=&  [ (\partial^2_t +  3 \dot\rho\, \partial_t + p^2) \delta^I_{J}
 + (V^I_{\,J}- e^{-3 \rho} \dot A^I_J ) ] u^J_{~\alpha,p}(t),
 \nonumber \\ \label{Eq:uIalpha}
\end{eqnarray}
we expand $\breve{\varphi}^I_1(x)$ as
\begin{eqnarray}
 \breve{\varphi}^I_1(x) = 
 \frac{1}{V_c^{\frac{1}{2}}} \sum_{{\sbm p}} \sum_{\alpha = 1}^D 
  \left\{ {u^I_{~\alpha,p}(t)\over M_{\rm pl}} 
 e^{i {\sbm p} \cdot {\sbm x}} a_{\alpha,{\sbm p}} +
 \mbox{h.c.}  \right\}, \label{Expansion:varphiI}
\end{eqnarray}
where the index 
$\alpha=1\cdots D$ is the label of the orthonormal basis.  
Making use of the Gram-Schmidt orthogonalization, the mode functions 
$\phi^I_{\alpha,{\sbm p}}\equiv 
e^{i{\sbm p}\cdot{\sbm x}}u^I_{\alpha,p}/M_{\rm pl} V_c^{1/2}$ 
are orthonormalized such that 
\begin{eqnarray}
(\phi_{\alpha,{\sbm p}},\phi_{\beta,{\sbm p}'})=
V_c \delta_{\alpha \beta} \delta_{\sbm{p} \sbm{p}'}~,
 \label{normalization}
\end{eqnarray}
is satisfied,  
where the Klein-Gordon inner product is defined by  
\begin{eqnarray}
 ( \phi,\,\psi  )= - i a^3 \int_{\Sigma}  \{\phi^I \partial_a
 \psi_I^{*}
  - \left( \partial_a \phi^I \right)
 \psi_I^*  \} d \Sigma^a~.  
\end{eqnarray}
In Eq.~(\ref{normalization}) 
the factor $V_c$ is necessary in order that 
the same functional form of the mode functions satisfies 
the natural orthonormal 
conditions in the continuum limit, $V_c\to\infty$. 
(See Appendix \ref{D and C}. ) 
Using the normalization conditions (\ref{normalization}), 
we find that the creation and
annihilation operators $a_{\alpha, {\sbm p}}^\dagger$ and
$a_{\alpha,{\sbm p}}$ satisfy the following commutation relations,
\begin{eqnarray}
 [ a_{\alpha,{\sbm p}},~a_{\beta, {\sbm p}'}^\dagger]
 =  ~\delta_{\alpha \beta} \delta_{{\sbm p} {\sbm p}'} ~.
 \label{commutation}
\end{eqnarray}
The initial vacuum state $\vert 0\rangle_a$ 
is annihilated by the operation of any annihilation 
operator: 
\begin{eqnarray*}
\! a_{\alpha, {\sbm p}} |0\rangle_a =0, \hspace{1cm} \mbox{for}~~
  ^\forall{\alpha} ~\mbox{and}~^\forall{\bm p} ~.   
\end{eqnarray*}

The mode function $u^I_{\alpha, {\sbm p}}(t)$ is normalized by
\begin{eqnarray}
 u^I_{\alpha,p}(t) \dot u_{I \beta,p}^{*}(t) -
 \dot u^I_{\alpha,p}(t) u^{*}_{I \beta,p} (t) = 
{i\over a^{3}(t)}\, \delta_{\alpha \beta}~.
 \label{Wronskian}
\end{eqnarray}
In the long wavelength limit, 
we obtain two real independent growing and decaying solutions as 
\begin{eqnarray}
 &&g^I_{\,\alpha,p}(t) = g^I_{\,\alpha}(t)\left[1+ 
          O \bigl((p/aH)^2 \bigr)\right]~,\cr
 &&d^I_{\,\alpha,p}(t) = d^I_{\,\alpha}(t)\left[1+ 
          O \bigl((p/aH)^2 \bigr)\right]~,
\label{gdvk}
\end{eqnarray}
and 
 $g^I_{~\alpha}(t)$ and $d^I_{~\alpha}(t)$ 
satisfy the normalization condition 
\begin{eqnarray}
 \dot g^I_{\,\alpha}(t) d_{I \beta}(t) -
 g^I_{\,\alpha}(t) \dot d_{I \beta} (t) = a^{-3}(t)\,\delta_{\alpha \beta}~.
\end{eqnarray}
$(g_\alpha,d_\beta)$ are the time
dependent solutions. In massless de Sitter approximation, they are given by
$ g^I_{\,\alpha}(t) \cong \delta^I_{\,\alpha}$ and 
$d^I_{\,\alpha}(t) \cong 1/(6H a^3) \delta^I_{\,\alpha}$, respectively.
Combining these two solutions, we can construct a mode function as 
\begin{eqnarray}
 u^I_{\alpha,k}(t) ={1\over c_\alpha(k)} g^I_{~\alpha,k}(t) 
 + i c_\alpha^*(k) d^I_{~\alpha, k}(t) , 
\label{formalvk}
\end{eqnarray}
with an arbitrary parameter $c_\alpha(k)$.
The squared amplitude of
$u^I_{\alpha,k}(t)$ gives the amplitude of the primordial
perturbations. It is common to set the initial state to the adiabatic
vacuum which is a natural state in the inflationary universe.
At the horizon crossing, where $k\approx aH$, the growing and decaying solutions 
should contribute to the positive frequency function $u^I_{\alpha,k}(t)$
to the same order unless the initial quantum state is very different from the 
adiabatic vacuum one. Assuming that the time variations of 
$g^I_{\, \alpha, k}$ and $a^3 H d^I_{\, \alpha, k}$ are not very fast 
after the horizon crossing time, i.e. 
$|g^I_{~\alpha, k}| /| d^I_{~\alpha, k}| \approx 
H a^{3+\delta_\alpha}$ with $\delta_\alpha \ll 1$, 
this requirement determines the order of magnitude of $c_\alpha(k)$ as
\begin{eqnarray}
 |c_\alpha(k)|=O\left(\sqrt{k^{3+\delta_\alpha} \over H^{2 + \delta_\alpha}} \right).
\label{formalvk3}
\end{eqnarray} 

Thanks to the local gauge conditions, 
as in the single field case discussed in our previous paper~\cite{Urakawa:2009my},
we can prove the regularity of the
IR fluctuations initially in the adiabatic direction, i.e. 
the tangential direction to the background trajectory. 
Looking at Eq.~(\ref{Eq:uIalpha}), we find that 
$\dot{\phi}^I / \dot{\rho} = d \phi^I / d \rho$  satisfies the mode
equation for the homogeneous mode $u^I_{~\alpha, 0}$. We choose one of 
the bases $u^I_{1, k}\approx g^I_{1,k}/c_1(k)$ so as to approach 
$d\phi^I / d \rho$ in the homogeneous limit $k\to 0$. 
Then, as we will show in \S \ref{Sketch}, the
modes with $\alpha=1$ no longer cause IR divergences. 
We give the other modes $u^I_{\bar{\alpha}, k} ~( \bar\alpha = 2,
\cdots D)$ so as to be orthogonal to
each other.

As we are considering the universe in a finite box, 
wave numbers ${\bm k}$ are discrete. 
Hence, unless we take the infinite volume limit, $V_c\to \infty$,  
the divergence is concentrated on the spatially homogeneous mode with $k=0$ in 
the above expression for the mode functions. 
To deal with this divergence 
in the $k=0$ mode, we regularize $c_{\baralpha}(0)$, 
introducing a small parameter $s_{\baralpha}$, as  
$$
c_{\baralpha}(0) \equiv  s_{\baralpha} / V_c^{\frac{1}{2}}~.
$$
Then, we obtain
\begin{eqnarray}
 u^I_{\baralpha, 0} (\tau) = \frac{V_c^{\frac{1}{2}}}{s_{\baralpha}}
  g^I_{~\baralpha,0}(t) + i \frac{s_{\baralpha}}{V_c^{\frac{1}{2}}}~  
  d^I_{~\baralpha,0}(t)~, 
 \label{Def:u0}
\end{eqnarray}
After we define appropriate observables, we take
the limit $s_{\baralpha} \rightarrow 0$ and $V_c \rightarrow \infty$. 

Giving the Wightman function $G^+_{IJ}(x,x')$ in Eq.~(\ref{Def:GR}) as
$G^+_{IJ}(x,x') = _a \langle 0|\breve{\varphi}_{I,1}(x) \breve{\varphi}_{J,1}(x') | 0 \rangle_a$,
we can expand the retarded Green function in terms of mode functions $u^I_{~\alpha}$ as 
\begin{eqnarray}
 \GR{I}{J}(x,~x') = -i \theta(t - t')~
\frac{1}{V_c} \sum_{{\sbm k}}
 ~e^{i {\sbm k} \cdot ({\sbm x} - {\sbm x}') } R^I_{\, J,k}(t, t'),\cr
 \label{GR mode}
\end{eqnarray}
where 
\begin{eqnarray}
 R^I_{\, J,k}(t, t') \equiv \sum_{\alpha = 1}^D \{  u^I_{\, \alpha,k}(t) u^*_{J \alpha, k}(t') 
- \mbox{c.c.}~\}. \label{Def:Rk}
\end{eqnarray}
Then, using the expressions in Eq.~(\ref{formalvk}),
we find that $R^I_{\, J,k}$ is regular even in the IR limit $k\to 0$. 

\subsection{IR vanishing smooth function}  \label{Sketch}
In this paper we do not consider the secular growth 
of the amplitude of perturbation due to the integration
for a long period of time. 
Namely, we consider the case 
that $t_i$ is set at a finite past from $t_f$. 
Deferring the detailed explanation to our succeeding paper, we give a
brief comment on the regularization of the secular growth in multi-field model in
Sec.\ref{conclusion}. In this paper we concentrate on the 
IR divergences originating from the momentum integration. 

The first part of our proof of IR regularity in multi-field model 
goes in parallel with the single field case~\cite{Urakawa:2009my}. 
In the single field model, the proof of IR regularity was quite simple
if we do not care about long time integration. 
However, multi-field extension turns out to be non-trivial even 
for this restricted case. 
To keep the simplicity of notation, we suppress 
the field indices $I$ and the labels
of modes $\alpha$ for a moment. 
As $\tilde\varphi(x)$ is composed of 
$\tilde \varphi_n~(n=1,2,3,\cdots)$, 
we make use of the mathematical induction to show
the regularity of all $\tilde\varphi_n$. 
$\tilde\varphi_n(x)$ is, by definition, $n$-th order in 
the interaction picture field $\breve \varphi_1$. 
Formally, we define $C[\tilde\varphi_n] 
(x; {\bm p}_1, \cdots, {\bm p}_n)$ by 
expanding $\tilde\varphi_n(x)$ as 
\begin{eqnarray}
\tilde\varphi_n(x)&=&
\left[ 
\prod_{j=1}^n \frac{1}{V^{\frac{1}{2}}_c} \sum_{\sbm{p}_j\ne 0}
 { a_{{\sbm p}_j} \over p_j^{{3}/{2}}} \right]
C^{(0)}[\tilde\varphi_n] (x; {\bm p}_1, \cdots, {\bm p}_n)\cr
&&+
a_0
\left[ 
\prod_{j=1}^{n-1} \frac{1}{V^{\frac{1}{2}}_c} \sum_{\sbm{p}_j\ne 0}
 { a_{{\sbm p}_j} \over p_j^{{3}/{2}}} \right]\cr
&&\qquad \times 
C^{(1)}
[\tilde\varphi_n] (x; {\bm p}_1, \cdots, {\bm p}_{n-1})
+\cdots~,
\label{Cexpansion}
\end{eqnarray}
where we have suppressed the terms containing creation operators. 
$C^{(j)}$ represents the coefficient of the term which contains $j$-th
order product 
of 0-mode operators $ a_{0}$. 
We also suppress this superscript ${(j)}$, for simplicity. 
The above expression is the result that we obtain after 
conducting all the integrations over the intermediate vertexes.
The momenta $\{{\bm p}_j\}$ in the argument of $C[\tilde\varphi_n]$
are those associated with the right most ends of the corresponding 
tree-shaped graph. 

A key ingredient of the first part of our proof is to show that
$C[\tilde\varphi_n] (x;{\bm p}_1, \cdots, {\bm p}_n)$ has the following properties, 
\begin{itemize}
\item
It is a smooth function with respect to $x$ 
for $^\forall p_j\equiv|{\bm p}_j| < a(t)\Lambda$, where 
$\Lambda$ is an UV momentum cutoff scale. 

\item
It vanishes when the long wavelength limit $p_j\to 0$ is 
taken for any momentum in its arguments.

\end{itemize}
If $C[\tilde\varphi_n]$ satisfies the properties mentioned above, (then
we say $C[\tilde\varphi_n]$ is an IR vanishing smooth function (IRVSF)), 
one can easily show that $n$-point functions 
$\langle \tilde\varphi(t_f,{\bm x}_1)\cdots \tilde\varphi(t_f,{\bm x}_n)
\rangle$ are free from IR divergences. 
When we take the expectation value of the product of $\tilde\varphi_j (j<n)$ 
in the form of Eq.~(\ref{Cexpansion}),  
we consider all the possible ways of pairing 
$a_{{\sbm p}_i}$ with $a_{{\sbm p}_j}^\dagger$. 
Then, each pair of $a_{{\sbm p}_i}$ and $a_{{\sbm p}_j}^\dagger$ is 
replaced with $\delta_{{\sbm p}_i, {\sbm p}_j}$.
One of the momentum integrations over ${\bm p}_i$ and ${\bm p}_j$ is 
performed to obtain an expression in the form 
\begin{eqnarray*}
 \int {d^3\!p_j \over (2\pi p_j)^3} C[\tilde\varphi_{n_1}](x_1;\cdots,{\bm
  p}_j,\cdots) 
C[\tilde\varphi_{n_2}](x_2;\cdots,{\bm p}_j,\cdots).  
\end{eqnarray*}
in the continuum limit. Here we note that 
$ V^{- \frac{1}{2}}_c \sum_{j=1} a_{{\sbm p}_j}$ 
should be replaced with
$ (2 \pi)^{-3} \int d^3 p_j \,a_{{\sbm p}_j}$ in the continuous
limit. (See Appendix \ref{D and C}.) 
The resulting momentum integration does not have IR divergences 
owing to the second property of $C[\tilde\varphi_{n}]$, i.e. 
$\lim_{\sbm{p}\to 0}C[\tilde\varphi_{n}](x;\cdots,{\bm
  p},\cdots) =0$.

Before we start the mathematical induction, let us note the following
properties of IRVSFs: 
\begin{description}
\item[Lemma]
If $C_1 (x;\{{\bm p}_j\})$ and $C_2 (x;\{{\bm q}_j\})$ 
are IRVSFs and there is no overlap between the list of momenta $\{{\bm p}_j\}$ 
and $\{{\bm q}_j\}$, then 
$\nabla_a C_1({x};\{{\bm p}_j\} )$,  
${\bm x}\, C_1(x;\{{\bm p}_j\})$, $\dot C_1(x;\{{\bm p}_j\})$,
	   $\triangle^{-1} C_1(x;\{{\bm p}_j\})$, 
$\hat{\bar W}_t C_1(x;\{{\bm p}_j\})$, 
$\int dt\, C_1(x;\{{\bm p}_j\})$, 
and $C_1(x;\{{\bm p}_j\}) \times C_2(x;\{{\bm q}_j\})$ 
are all IRVSFs.
\end{description}

Now, let us prove that $C[\tilde\varphi_n]$ 
is IRVSF by induction if $C[\tilde\varphi_1]$ is so.  
The $n$-th order perturbation is obtained by 
\begin{eqnarray}
\tilde\varphi_n
&\! =&\! \hat{\bar W}_t\!\int^t\! dt'\! 
\int\! d^3\!x' a^3(t') G_R(x,x')
W_{t'}(x') \Gamma_n(x')~. \nonumber \\
\label{WGWG}
\end{eqnarray} 
$W_{t'}(x') \Gamma_n(x')$ 
is constructed from lower order perturbations 
$\delta {\tilde N}_j$, ${\tilde \chi}_j$, $f_j$, $B_{\perp j}$
and $\tilde\varphi_j$ with $j<n$ using the operations listed in 
the above Lemma. Furthermore, from Eqs.~(\ref{Eq:Ni}), (\ref{Eq:chii})
and (\ref{Bfeq}), we find that 
$\delta {\tilde N}_j$, ${\tilde \chi}_j$ and $f_j$ are 
all constructed from $\tilde\varphi_l$ with $l \leq j$ by the operations listed there, too. 
Hence, $C[W_{t'} \Gamma_n]$, 
the expansion coefficient of $W_{t'}(x') \Gamma_n(x')$ 
analogous to $C[\tilde \varphi_n]$ 
in Eq.~(\ref{Cexpansion}), is also an IRVSF. 
Since the Fourier mode of the retarded Green function described by
 Eq.~(\ref{Def:Rk}) is regular in the IR limit, its Fourier transform 
$G_R(x,x')$ should be regular, too. (Regularity in UV is assumed to be
guaranteed by an appropriate UV renormalization.) 
Since the integration volume of ${\bm x}'$ is finite, 
the integral of a product of regular functions $\int\! d^3\!x' a^3(t') G_R(x,x')
W_{t'}(x') \Gamma_n(x')$ should be finite, and hence it is IRVSF. 
Since the operation $\hat{\bar W}_t$ preserves the properties of 
IRVSF,  
$\tilde\varphi_n=\hat{\bar W}_t
\hat G_R (W_{t'} \Gamma_n)$ is also found to be IRVSF.


Now our concern is whether the first step of induction is 
true or not. Namely, we examine if $C[\tilde\varphi_1]$ is IRVSF or not. 
Utilizing the residual gauge degrees of freedom, we fix the local
average of the adiabatic mode $\Asigma=e_I\tilde\varphi^I$. 
Then,
IR modes in this direction are controlled to be free from 
divergences, but the modes in the other directions are not. 
The interaction picture field appears only in the projected form 
$
\hat{\bar{W}}^{I}_{\,J} \breve{\varphi}^J_{1}
$, which 
can be expanded by the mode function $u^I_{~\alpha,k}$ as 
\begin{eqnarray}
 \hat{\bar{W}}^{I}_{\,J} \breve{\varphi}^J_{1}(x) 
 &=&  \frac{1}{V_c^{\frac{1}{2}}} 
 \sum_{\alpha,{\sbm p}} \left[e^{i{\sbm p}\cdot{\sbm x}} {\cal G}^I_{~J} 
 - \frac{W_{t, - {\sbm p}}}{W_{t,0}} e^I e_J \right] {u^J_{
  \alpha, p}(t) \over M_{\rm pl} }
          a_{\alpha, {\sbm p}}  \cr
 && \qquad \qquad  +\{\rm h.c.\}, 
\end{eqnarray}
where 
\begin{eqnarray}
W_{t,-{\sbm p}}\equiv\int d^3\!x \,e^{i{\sbm p}\cdot{\sbm
x}}\, W_t({\bm x})~, 
\end{eqnarray} 
and we note that $W_{t,{\sbm 0}}=\int d^3\! x\, W_{t}({\bm x})=L_t^3$.
To make it easy to take the limit $V_c \rightarrow\infty$, we define the Fourier
mode of the window function in a different manner from those of 
fluctuations. (See Appendix \ref{D and C}.)
Hence, we have the coefficient for $a_{\alpha, {\sbm p}}$ as
\begin{equation}
 C^{(0)}_{\alpha}[\hat{\bar{W}}^{I}_{\,J} \breve{\varphi}^J_{1}
   ](x,{\bm p})=
  \left[e^{i{\sbm p}\cdot{\sbm x}} {\delta}^I_{~J} 
 - \frac{W_{t, - {\sbm p}}}{W_{t,0}} e^I e_J \right] {p^{\frac{3}{2}}u^J_{
  \alpha, p}(t) \over M_{\rm pl} } .  
\label{poneexp}
\end{equation}
We have chosen the adiabatic mode ($\alpha = 1$) so as to be 
tangential to the background trajectory in $p \rightarrow 0$ limit, i.e., 
$g^I_{~1}(t) \propto e^I$. 
Then, multiplying
$p^{\frac{3}{2}} u^I_{~1,p}(t) \cong p^{- \frac{\delta_1}{2}}$ 
by the factor 
$[e^{i{\sbm p}\cdot{\sbm x}}-W_{t,-{\sbm p}} / W_0 ]$,~ 
$C^{(0)}_1 [\hat{\bar{W}}^{I}_{\,J} \breve{\varphi}^J_{1}
   ](x,{\bm p})$ vanishes in  this limit.  Therefore 
$ C^{(0)}_{\alpha}[\hat{\bar{W}}^{I}_{\,J} \breve{\varphi}^J_{1}
   ](x,{\bm p})
$ vanishes 
in the limit $p\to 0$.  
Thus, we find that 
$ C^{(0)}_{\alpha}[\hat{\bar{W}}^{I}_{\,J} \breve{\varphi}^J_{1}
   ](x,{\bm p})$
is an IRVSF. 
However, the factor 
$\left[e^{i{\sbm p}\cdot{\sbm x}} {\delta}^I_{~J} 
 -  W_{t, - {\sbm p}} / W_{t,0}  e^I e_J \right]$ does not suppress
isocurvature fluctuations $\bS^I$, which is pointing 
orthogonal direction to the background trajectory. 
When one of the basis with $\bar \alpha \ \ne 1$ has non-negative
value of $\delta_{\bar \alpha}$, the IR contribution of such a mode 
diverges and 
$ C^{(0)}_{\alpha}[\hat{\bar{W}}^{I}_{\,J}\breve{\varphi}^J_{1}
   ](x,{\bm p})$ is not IRVSF.
In this case $n$-point functions of $\tilde{O}(x)=\{\Asigma,\tS^I \}$
actually diverge. 

The case with ${\bm p}=0$ goes in a similar manner.
The coefficient for $a_{\alpha, 0}$ of $\tilde\varphi^I_1$ is 
given by 
\begin{eqnarray}
 C^{(1)}_{\alpha}[\hat{\bar{W}}^{I}_{\,J}\breve{\varphi}^J_{1}](x) 
  &\!=&\!  
  \left[{\delta}^I_{~J} - e^I e_J \right] {g^J_{
  \alpha, 0}(t) \over s_\alpha M_{\rm pl} }~.
\label{zerooneexp}
\end{eqnarray}
where we take the limit $V_c\to \infty$ using Eq.~(\ref{Def:u0}).  
This expression also vanishes for the adiabatic perturbation, but 
it does not for the isocurvature perturbation. Then, the contribution 
from the isocurvature perturbation diverges when the limit 
$s_\alpha\to 0$ is taken. 

The above divergences arise only 
in the multi-field model. 
In the rest of this section, we discuss the
regularization of this divergence.
Even when $\delta_{ \bar \alpha}$ 
is negative, the following discussion is still relevant. 
For $\delta_{ \bar \alpha} <0$ there is no IR divergence, but IR contribution  
can be large if $|\delta_{ \bar \alpha} |\ll 1$. 
If one can show that 
$n$-point functions for $\tilde{O}(x)$ are free from the IR divergence 
for $\delta_\balpha >0$, it also implies the absence of enhanced IR contribution
for $\delta_\balpha <0$.


\subsection{Squeezed wave packet} \label{Squeezed}
For the later use, we transform the mode functions 
$\{ u^I_{~\alpha, p} e^{i {\sbm p} \cdot {\sbm x}}\}$ to another ones
suitable for discussing the effect of inserting projection operator. 
Transformation proceeds in two steps. 
Deferring the detailed explanations to Appendix \ref{Sec:Bogoliubov},
here we just give a brief sketch of the transformations.
At the first step 
the new basis mode functions $\{v^I_{~ \baralpha, {\sbm p}}\}$ 
for ${\bm p}\ne 0$ are
arranged so that the leading term in the long wavelength limit is
cancelled. 
While, the IR divergent contributions are
localized to a single mode with $p=0$. At the second step of
transformation, without changing the mode function for ${\bm p} \neq 0$,
we introduce another mode function for $p=0$ mode which 
naturally defines wave packets with a finite width
even in the limit $s_{\bar \alpha} \rightarrow 0$ and 
$V_c \rightarrow \infty$. In this limit, 
$\breve{\varphi}^I_1$ can be expanded as
\begin{eqnarray}
\breve{\varphi}^I_1(x) \!&=\!& \sum_{\baralpha} \left\{ {\bar{v}^I_{~\baralpha,0} \over \Mp}
  \bara_{\baralpha,0} + 
 \int_{\sbm{p} \ne 0}\! \frac{d^3 {\bm p}}{(2 \pi)^{\frac{3}{2}}} 
  {\bar{v}^I_{~\baralpha, \sbm{p}}(t)\over M_{\rm pl}}
  \bara_{\alpha,{\sbm p}}  \right\}\nonumber\\
 &&\qquad\qquad + (\mbox{h.c.})~,
\label{Def:varphis}
\end{eqnarray}
where
\begin{eqnarray}
 \bar{v}^I_{\,\baralpha,0} (x) \!&=\!& 
 H_f g^I_{\,\baralpha,0}(t) 
 + \frac{i}{H_f} \int_{\sbm{p} \ne 0} \frac{d^3 \bm{p}}{(2 \pi)^3}
  \frac{W_{\sbm p}}{W_0} 
  d^I_{\,\baralpha,p}(t) e^{i \sbm{p} \cdot \sbm{x}}~, \nonumber \\ 
 \label{v0kb}\\
 \bar{v}^I_{\, \baralpha,\sbm{p}} (x) \!&=\!& u^I_{\, \baralpha, p}(t) e^{i \sbm{p} \cdot \sbm{x}} 
    - \frac{W_{- \sbm{p}} }{W_0} {g^I_{\,\balpha,0} (t) \over
    c_{\baralpha}(p)}~. 
 \label{vpkb}
\end{eqnarray}
Now, we introduce the coefficients 
$\bar C^{k}_{\{\alpha\}}[\tilde\varphi^I_{n}]$ 
analogous to $C^{k}_{\{\alpha\}}[\tilde\varphi^I_{n}]$ 
for the creation and annihilation operators 
$\bara_{\baralpha,\sbm{p}}^\dagger$ and 
$\bara_{\baralpha,\sbm{p}}$.
Then, in the continuum limit, $\tilde\varphi^I_{n}$ is 
expanded as 
\begin{eqnarray}
\tilde\varphi^I_n(x)&=&\sum_{k=0}^n
\left[
  \prod_{l =1}^{k}
 \bara_{\alpha_{n-k+l},0}\right]
\left[ \prod_{j=1}^{n-k} 
 \int 
 { d^3 \bm{p}_j \over (2 \pi p_j)^{\frac{3}{2}}} \bara_{\alpha_j, {\sbm
 p}_j} \right] \cr
&&\times
\bar C^{(k)}_{\{\alpha\}}[\tilde\varphi^I_n] (x; {\bm p}_1, \cdots, {\bm p}_{n-k})\cr
&&\qquad \qquad \qquad
+\cdots~. \label{Def:CI}
\end{eqnarray}
Since the induction with respect to $n$ proceeds as before, 
we can say that $\bar C^{(k)}_{\{\alpha\}}[\tilde\varphi^I_n]$ is IRVSF
if the coefficients of the first order 
variables $ \hat{\bar{W}}^{I}_{\,J}\breve{\varphi}^J_{1}(x)$ 
are IRVSFs. 

In the same way as Eq.~(\ref{poneexp}), 
the coefficient for $\bara_{\alpha, {\sbm p}}$ is obtained as 
\begin{eqnarray}
&& \hspace{-10mm}
  \bar C^{(0)}_{\alpha}[\hat{\bar{W}}^{I}_{\,J}\breve{\varphi}^J_{1}
  ](x,{\bm p})\cr
&& =
  \left[e^{i{\sbm p}\cdot{\sbm x}} {\delta}^I_{~J} 
 - \frac{W_{t, - {\sbm p}}}{W_{t,0}} e^I e_J \right]
  {p^{\frac{3}{2}}
\bar v^J_{\alpha, p}(t) \over M_{\rm pl} }~.  \label{barC0Wvarphi1}
\end{eqnarray}
From Eq.~(\ref{vpkb}), $p^{3/2}\bar v^J_{\balpha, p}$ vanishes 
in the limit $p\to 0$. The adiabatic component also vanishes 
owing to the projection $\hat{\bar W}^I_{\,J}$. Hence these coefficients are IRVSFs. 
While, the coefficient for $\bara_{\alpha,0}$ is given by 
\begin{eqnarray}
 C^{(1)}_{\alpha}[\hat{\bar{W}}^{I}_{\,J}\breve{\varphi}^J_{1}](x) 
  &\!=&\!  
  \left[{\delta}^I_{~J} - e^I e_J \right] {\bar v^J_{
  \alpha, 0}(t) \over M_{\rm pl} }~, 
\end{eqnarray}
which is finite from Eq.~(\ref{v0kb}) in contrast to the 
previous case in Eq.~(\ref{zerooneexp}). Therefore it is also IRVSF.
As we find that all the coefficients of $\hat{\bar W}^I_{\,J} \breve\varphi^J_1$ 
are IRVSFs or just a regular function independent of ${\bm p}$, 
all higher order coefficients
$\bar C_{\{\baralpha\}}^{(k)}[\tilde{\varphi}^I_n]$ are proven
to be IRVSFs by induction. 

We should emphasize that 
even if the coefficients $\bar C^{(k)}_{\{\alpha\}}[\tilde\varphi^I_n]$ 
are IRVSFs, this does not
imply the regularity of $n$-point functions of $\tilde{O}$ 
for the initial adiabatic vacuum state $|0\rangle_a$. 
In the following discussion we use an expression for 
the adiabatic vacuum state 
$|\,0\,\rangle_a$ expanded in terms of the coherent 
states $|\beta\rangle_{\bara}$ associated with $\bara_{\baralpha,0}$. 
The coherent state satisfies
\begin{eqnarray*} 
 && \bara_{\balpha,\sbm {p}} |\beta \rangle_{\bara} =
  \beta_{\balpha} |\beta \rangle_{\bara}~, \cr
  && \bara_{1,\sbm {p}} |\beta \rangle_{\bara}  =0~.
\end{eqnarray*}
As shown in Appendix \ref{Sec:Bogoliubov}, the original vacuum state
$|\,0\, \rangle_a $  can be expressed as
\begin{eqnarray}
 |\,0\, \rangle_a 
 &=& \prod_{\balpha=2}^D \left[ \int^{\infty}_{-\infty} 
     d\beta_{\balpha}\, E_{\balpha}(\beta_{\balpha}) \right] \,|\beta \rangle_{\bara}~, 
  \label{BD by cs}
\end{eqnarray}
where the coefficient $E_{\balpha}({\beta_{\balpha}})$ approaches to
\begin{eqnarray}
 E_{\balpha}(\beta_{\balpha}) &\to &
\sqrt{s_{\balpha} H_f \over \pi}e^{-(s_{\balpha} H_f \beta_{\balpha})^2},
\label{csdef}
\end{eqnarray}
in the limit of $s_{\balpha} \to 0$.
The non-vanishing support for $E_{\balpha} $ extends to infinitely large
$|\beta_{\baralpha}|$ in this limit. 
This is a consequence of the fact that 
the wave function in the adiabatic vacuum 
is highly squeezed in the direction corresponding to
$\bara_{\baralpha,0}$. 

Using Eq.~(\ref{BD by cs}), we can expand the 
$n$-point function of the ``observables''
$\tilde{O} (t_f, \bm{x})$ as 
\begin{eqnarray}
 && \hspace{-10mm}
  \langle ~\proj \tilde{O} (t_f, \bm{x}_1) \tilde{O}
  (t_f, \bm{x}_2)  \cdots \tilde{O} (t_f, \bm{x}_n)
  ~\rangle
 \nonumber \\ 
 &=& \prod_{\balpha=2}^D \left[ \int^{\infty}_{- \infty} d
			   \beta_{\balpha}  \int^{\infty}_{-
			   \infty} d \gamma_{\balpha}  
   \,E_{\balpha} (\beta_{\balpha}) E_{\balpha} (\gamma_{\balpha}) \right]
 \nonumber \\ && \qquad
 \times\, 
{}_{\bara} \langle \beta|~\proj~\tilde{O} (t_f,
 \bm{x}_1)
  \cdots \tilde{O} (t_f, \bm{x}_n)  ~  |\gamma
  \rangle_{\bara}^{\rm conn}. \nonumber \\
\label{Exp:npoint}
\end{eqnarray}
where, to clarify that we should sum up only the connected graphs, we have
added the suffix ``conn''.

One remark is in order in 
computing the expectation value of the product of
$\{\bara_{\alpha, \sbm{p}} \}$ and $\{\bara^{\dagger}_{\alpha,\sbm{p}} \}$. 
Basically pairs between $\bara_{\alpha,\sbm{p}}$ and $\bara^\dagger_{\alpha',
\sbm{p}'}$ are replaced with 
$\delta (\bm {p} - \bm {p}') \delta_{\alpha, \alpha'}$ 
except for the case with $\bm{p} = 0$. 
After the replacement, only the operators $\{\bara_{\baralpha\,0}\}$ and 
$\{\bara^{\dagger}_{\baralpha,0} \}$ are left on the right hand side of Eq.~(\ref{Exp:npoint}).
The expectation value of the product of these operators 
becomes
the summation of the commutators and their normal ordered products. 
Since the operators $\bara_{\baralpha,0}$ 
and $\bara_{\baralpha,0}^\dagger$ are not annihilated by 
the coherent state, the expectation value of the normal ordered products 
composed of these zero-mode operators are non-vanishing. 
The annihilation operators $\bara_{\baralpha,0}$ 
in the normal ordered parts acting on the coherent state
$|\gamma \rangle_{\bara}$ produces the factor $\gamma_{\baralpha}$, while 
the creation operator $\bara_{\baralpha,0}^\dagger$ acting on 
${}_{\bara}\langle\beta\vert$ 
produces $\beta_{\baralpha}$. 
In other words, $\bara_{\baralpha,0}$ and 
$\bara_{\baralpha,0}^\dagger$ are replaced either with a commutator by making a pair 
or with $\gamma_{\baralpha}$ and $\beta_{\baralpha}$, respectively. 
These two exclusive possibilities can be concisely 
expressed by the replacement
\begin{eqnarray}
 \bara_{\baralpha,0}~\rightarrow~\gamma_{\baralpha} +
  \bara_{\baralpha,0}^{(q)}~, ~~
 \bara^{\dagger}_{\baralpha, 0}~\rightarrow~\beta_{\baralpha} 
  + \bara_{\baralpha,0}^{(q)\dagger}~, 
 \label{replace b0}
\end{eqnarray}
where $\bara_{\baralpha,0}^{(q)}$ and $\bara_{\baralpha,0}^{(q)\dagger}$ 
satisfy the same commutation relation as
$\bara_{\baralpha,0}$ and $\bara_{\baralpha,0}^\dagger$, and they annihilate the coherent state
$|\gamma \rangle_{\bara}$ and
${}_{\bara}\langle\beta\vert$, respectively. 

Now it will be obvious that $n$-point function evaluated for the 
coherent states,  
\begin{equation}
{}_{\bara} \langle \beta|~\tilde{O} (t_f,
 \bm{x}_1)
  \cdots \tilde{O} (t_f, \bm{x}_n)  ~  |\gamma
  \rangle_{\bara}^{\rm conn}. \nonumber 
\end{equation}
is finite. 
To show its regularity, 
the insertion of the projection operator $\proj$ is unnecessary. 
We were focusing on $\bar C^{(j)}_{\{\alpha\}}[\tilde\varphi^I_n]$, 
which is the coefficients of the annihilation operators, 
but all the other coefficients of the mixture of $\{ \bara_{\alpha, \sbm {p}}\}$ 
and $\{ \bara_{\alpha, \sbm {p}}^\dagger\}$ are 
shown to be IRVSFs in the same
manner. The effect of coherent state is taken care by the replacements
(\ref{replace b0}). 
Therefore, the expectation
value for fixed values of $\beta_{\baralpha}$ and $\gamma_{\baralpha}$
is regular.

\subsection{Role of projection operator} \label{Projection}
Without the projection operator $\proj$, the $n$-point functions of
$\tilde{O}(x)$ for the adiabatic vacuum diverge, although the
expectation values for the coherent states were proven to be  
finite in the preceding subsection. 
The divergences for the adiabatic vacuum appear in the 
integration over $\beta_{\baralpha}$ and $\gamma_{\baralpha}$. 
The limiting behavior of 
$E_{\baralpha}(\beta_{\baralpha})$ for $s_\alpha\to 0$ 
given in Eq.~(\ref{csdef}) tells that this factor 
does not restrict the effective range of 
these integrations in the limit $s_{\baralpha} \to 0$. 
Therefore, integrating over $\beta_{\baralpha}$ and $\gamma_{\baralpha}$ 
without the insertion of $\proj$, the $n$-point function 
for the adiabatic vacuum diverges. This result is as expected 
since the basis transformation does not change the final result for the
$n$-point function. 

In order to remedy these divergences, we need the
insertion of the projection operator.  
The projection operator $\proj$ takes care of the effect of quantum
decoherence, removing the contamination from the other parallel worlds. 
We will see that the insertion of $\proj$
makes the effective range of integration for 
$\beta_{\baralpha}$ and  $\gamma_{\baralpha}$ finite. 

In the same way as $\tilde{O}(t_f, \bm{x})$,  
we expand $\proj$, which is the functional of
$\{\bS^\balpha(t_f) \}$, in terms of $\{\bara_{\alpha, \sbm{p}}\}$ and
$\{\bara^\dagger_{\alpha, \sbm{p}}\}$. 
Expanding $\bS^{\balpha}(t_f)$ as
\begin{eqnarray}
 \bS^\balpha(t_f) = \bS^\balpha_{1}(t_f) 
  + \frac{1}{2} \bS^\balpha_{2}(t_f) 
+ \cdots,  
\end{eqnarray}
we focus on the coefficient of $\bar a_{\balpha,0}+\bar a^\dag_{\balpha,0}$
in the leading term $\bS^\balpha_1(t_f)
\equiv e^\balpha_I(\hat{W}_{t_f}\tilde\varphi^I_1+B^I_{\perp 1})$. 
We decompose the terms which contain 
$\bar a_{\balpha,0}$ and $\bar a^\dag_{\balpha,0}$ 
into two pieces proportional to 
$(\bar a_{\balpha,0}+\bar a^\dag_{\balpha,0})$
and 
$(\bar a_{\balpha,0}-\bar a^\dag_{\balpha,0})$. 
Then, we can see that the former does not contain contribution from
$B^I_{\perp 1}$ as follows. 
$B^I_{\perp 1}$ is determined  by solving
Eq.~(\ref{Bfeq}), which is sourced by 
$e_I \hat W_t\tilde\varphi^I_1$ on the right hand side. 
The coefficient of $\bar a_{\balpha,0}+\bar a^\dag_{\balpha,0}$ in 
$\hat W_t\tilde\varphi^I_1$ is given by 
\begin{eqnarray}
&& \hspace{-1cm}
 {\hat W_t\over 2}(\bar v^I_{\,\balpha,0}+\bar v^{*I}_{\,\balpha,0})
 =H_f g^I_{\balpha,0}
    +{i\over 2H_f}\int_{{\sbm p}\ne 0}{d^3{\bm p}\over(2\pi)^3}
\cr &&\qquad\qquad\qquad \times
\biggl({W_{\sbm p} W_{t,-{\sbm p}}
  \over W_{0} W_{t,0}} d^I_{\,\balpha, p}
  - (\mbox{c.c.})\biggr)~. 
\end{eqnarray}
Using the relation $W_{t,-{\sbm p}}=W^*_{t,{\sbm p}}$, which is derived
from the reality condition of $W_t(x)$, the second term on the 
right hand side vanishes. Contraction with $e_I$ erases the first term,
too. Since the source term of the equation for $B^I_{\perp 1}$ does not contain 
$(\bar a_{\balpha,0}+\bar a^\dag_{\balpha,0})$, 
$B^I_{\perp 1}$ does not, either. 

Using the replacement (\ref{replace b0}), we divide 
$(\bar a_{\balpha,0}+\bar a^\dag_{\balpha,0})$ into 
$(\beta_{\balpha}+\gamma_{\balpha})$ 
and $(\bar a^{(q)}_{\balpha,0}+\bar
a^{(q)\dag}_{\balpha,0})$. 
Separating the coefficient of $(\beta_{\balpha}+\gamma_{\balpha})$, 
$\bS^\balpha_1(t_f)$ is expressed as 
\begin{eqnarray}
  \bS^\balpha_1(t_f)=
    \Delta^\balpha_{~\bbeta} (\beta^{\bbeta}+\gamma^{\bbeta})+ 
   \delta\bS^\balpha_1(t_f), 
\end{eqnarray}
where 
\begin{eqnarray}
\Delta^\balpha_{~\bbeta} \equiv \frac{H_f}{\Mp} 
   e^\balpha_I g^I_{\,\baralpha,0}(t_f),  
\end{eqnarray}
and summation over repeated index $\bbeta$ is understood. 

Inserting the above expression for $\bS^\balpha(t_f)$ 
into $\proj$ given in
(\ref{gaussianproj}), the observable $n$-point function
(\ref{Exp:npoint}) is recast into
\begin{eqnarray}
 && \hspace{-5mm}\langle ~\proj\,\tilde{O} (t_f, \bm{x}_1) \tilde{O}
  (t_f, \bm{x}_2)  \cdots \tilde{O} (t_f, \bm{x}_n)  ~\rangle
 \nonumber \\ 
 &=&  
 \!\prod_{\bar\epsilon=2}^D\left[
\int^{\infty}_{- \infty} d \beta^{\bar\epsilon}
   \! \int^{\infty}_{-\infty} d \gamma^{\bar\epsilon}
   \,E (\beta^{\bar\epsilon}) 
 E (\gamma^{\bar\epsilon}) \right]\nonumber \\ 
&&
 \exp\left[-{\cov^{-1}_{\balpha\bbeta}
  \Delta^\balpha_{\,\bgamma} (\gamma^\bgamma + \beta^\bgamma)
  \Delta^\bbeta_{\,\bdelta} (\gamma^\bdelta + \beta^\bdelta) \over 2}
  \right] \nonumber \\ 
&&
 _{\bara} 
  \langle \beta| 
\exp\left[
  - \frac{\cov^{-1}_{\balpha\bbeta}\left\{
   2 \Delta^\balpha_{\,\bgamma} (\gamma^\bgamma + \beta^\bgamma)
            \delta\bS^\bbeta
  +  \delta\bS^\balpha \delta\bS^\bbeta \right\}}{2} 
\right] \nonumber \\ && \quad \qquad \times
\tilde{O} (t_f, \bm{x}_1) \tilde{O} (t_f,\bm{x}_2)
  \cdots \tilde{O} (t_f, \bm{x}_n)    | \gamma
  \rangle_{\bara}^{\sc conn}~, \nonumber \\
\label{POO}
\end{eqnarray}
where 
\begin{eqnarray*}
 \delta \bS^\balpha\equiv \delta\bS^\balpha_1(t_f) 
+{1\over 2}\bS^\balpha_2(t_f)+\cdots~. 
\end{eqnarray*}

Owing to the first exponential factor, 
the contribution from the integration region with 
$|\Delta^\balpha_{\,\bbeta}(\gamma^\bbeta + \beta^\bbeta)| \gg \sigma$ 
is exponentially suppressed. 
Since the
inner product between the coherent states gives
\begin{eqnarray*}
{}_{\bara}\!
 \left\langle \beta\, \right\vert \left. 
    \gamma \right\rangle_{\bara} 
  = \prod_{\balpha=2}^D
 \exp\left[- \frac{(\beta^\balpha - \gamma^\balpha )^2}{2}\right],
\end{eqnarray*}
(see Eq.~(\ref{def coherentap})), the 
contribution from the region with
$|\beta^\balpha - \gamma^\balpha| \gg 1$ is 
also exponentially suppressed. 
The directions of these two suppression are orthogonal, 
and hence the effective integration area is restricted to 
a finite region 
\begin{equation}
| \beta^\balpha | ,~|\gamma^\balpha | 
  \alt \max\left(1,{\sigma \over
\Delta }\right), 
\label{deltaconst}
\end{equation}
where $\Delta$ and  $\sigma$ are the typical 
amplitudes of the eigenvalues of $\Delta^\balpha_{\,\bbeta}$ 
and the square of the eigenvalues 
of $\cov^\balpha_{\,\bbeta}$, 
respectively. 
Our discussion up to here ensures the finiteness 
of the effective range of the Gaussian integrations 
over $\beta^\balpha$ and $\gamma^\balpha$.  
This proves the IR regularity
of the $n$-point function of the local fluctuation
$ \{ \tilde{O}(x) \}$ with the projection $\proj$ 
at each order of loop expansion even if 
the initial state is set to 
the adiabatic vacuum state. 

Now the remaining task is to examine 
if the perturbative expansion is still reliable 
after all the changes that we made. 
For a sufficiently wide projection operator $\proj$ , 
the expected amplitude of 
$|\tilde \varphi^{I}|$ is dominated by 
the contribution from $\beta^\balpha$ and $\gamma^\balpha$, 
which is $O(\sigma)$. 
Hence, the validity of the perturbative expansion requires 
\begin{eqnarray}
 \sigma \ll 1. 
\label{sigmaupper}
\end{eqnarray}
The next question is whether one can safely expand 
the second exponential factor in (\ref{POO}), 
\begin{eqnarray*}
 e^{
  - \frac{\cov^{-1}_{\balpha\bbeta}\left\{
   2 \Delta^\balpha_{\,\bgamma} (\gamma^\bgamma + \beta^\bgamma)
            \delta\bS^\bbeta
  +  \delta\bS^\balpha \delta\bS^\bbeta \right\}}{2} 
},
\end{eqnarray*}
coming from $\proj$. 
Using the conditions (\ref{deltaconst}), 
we find that this expansion converges 
if 
$$
|\delta\bS^\balpha|\ll \sigma
$$ 
is satisfied. On the other hand, the amplitude of 
$|\delta\bS^\balpha|$ is 
estimated to be given by the linear order contribution 
$\Delta$. 
Therefore the necessary condition is 
\begin{eqnarray}
\Delta \ll \sigma.
\label{Delta sigma}
\end{eqnarray}
Since $\Delta \ll 1$, we can choose $\sigma$ such that satisfies 
(\ref{sigmaupper}) and (\ref{Delta sigma}) simultaneously.

\section{Conclusion}  \label{conclusion}
In the present paper, we have proposed one solution to the IR divergence problem in
multi-field models. 
We discuss $n$-point functions for the local
perturbations variables, $\tilde{O}(x)$, defined by the deviations from 
the local average values with an additional gauge condition that fixes 
one of the residual gauge degrees of freedom remaining in the usual flat gauge. 
Even if we consider these local perturbative variables, 
when plural fields have a scale invariant or red spectrum, we encounter
IR divergences. This is because the effects of quantum decoherence 
are not taken into account yet;
the $n$-point function for $\tilde{O}(x)$ is affected by the contaminations from
other uncorrelated worlds. To remove the contaminations, we have inserted the
projection operator $\proj$ which projects the final quantum state 
to a wave packet with a sharp peak in the space of the local average
values of the fields, $\{ \hat{W}_{t_f} {\cal S}_I \}$. 
Here, we give an intuitive way to understand how the insertion of the
projection operator regularizes the IR corrections. When 
plural fields contribute to IR divergences, the wave
function corresponding to the initial adiabatic vacuum is highly 
squeezed in the corresponding directions in the space of $\{ \hat{W}_{t_f} {\cal S}_I \}$.  
However, only a part of the squeezed wave function does
contribute to a decohered wave packet, which represents our world. 
Introducing the projection 
operator, we have taken into account the restriction of the wave
function to the non-vanishing support of the decohered wave packet. 
This restriction is recast into the exponential factor 
in Eq.~(\ref{POO}), by which
the non-vanishing support of $(\beta, \gamma)$ becomes a finite
region. This assures the regularity of the observable $n$-point
functions. 

The ``observable'' n-point function (\ref{obs npointfn})
depends on the parameter $\sigma$ that we introduced to incorporate the
decoherence effect without discussing the detailed process. It also
depends on the size of the observable region. These dependencies may
disappear when we compute the actual observables, taking into account the
decoherence process appropriately. Definitely, to predict the accurate value
of observables, further study is necessary. We leave this issue for our
future work.

Here, fixing the temporal coordinate on each vertex, we showed the
regularity of the integration over the spatial coordinates. Hence, we
cannot deny the possibility that the temporal integral makes the $n$-point
function diverge when we send the initial time $t_i$ to a distant
past. In single-field models, we showed, by considering the local
fluctuations, that we can suppress the secular growth which appears from the
temporal integration, unless very higher order perturbation is considered.   
We expect that in the multi-field model, a similar suppression appears
from the inserted projection operator. The reason is as follows. If the
non-vanishing contributions from the distant past makes the amplitude of 
$\tilde{O}(t_f,~\bm{x})$ significantly large, they must increase also the
amplitude of the local average $\{ \hat{W}_{t_f} {\cal S}_I \}$ to
a large extent. However, the decohered wave packet we pick up has the
bounded average values of $\{ \hat{W}_{t_f} {\cal S}_I \}$. Therefore,
we expect that as long as we compute the observable $n$-point function evaluated for
a decohered wave packet, the contributions from the distant past are
suppressed and the temporal integration converges. We also defer the examination of
our optimistic expectation to a future work.

In general, the IR divergences originate from massless
(or quasi-massless) fields with non-linear interactions in the (quasi-)
de Sitter universe~\cite{Onemli:2002hr, Brunier:2004sb,Prokopec:2007ak,
Tsamis:1996qm, Tsamis:1996qq, Garriga:2007zk, Tsamis:2007is}. 
In~\cite{Onemli:2002hr, Brunier:2004sb,Prokopec:2007ak}, 
using the prescription with an IR cutoff, the logarithmic amplification is discussed
for a massless test scalar field $\phi$ with a quadratic interaction term. 
We can discuss the regularization of the IR corrections from a test
field in a similar manner to the entropy perturbations discussed in this
paper. 
In~\cite{Tsamis:1996qm, Tsamis:1996qq}, the effects of IR gravitons which grow
logarithmically are argued to screen the cosmological constant. However,
the Hubble parameter defined in \cite{Tsamis:1996qm, Tsamis:1996qq} is
gauge dependent and in other definition we do not encounter the
screening \cite{Garriga:2007zk}. 
Thus, this problem is still controversial~\cite{Garriga:2007zk, Tsamis:2007is}.
We would like to apply our prescription to these issues in our future
work, too. 

\begin{acknowledgments}
 YU would like to thank Kei-ichi Maeda for his continuously
 encouragement. YU is supported by the JSPS under Contact No. 19-720.
TT is supported by Monbukagakusho Grant-in-Aid for
Scientific Research Nos. 19540285 and 21244033. This
work is also supported in part by the Global COE Program ``The
Next Generation of Physics, Spun from Universality and
Emergence'' from the Ministry of Education, Culture,
Sports, Science and Technology (MEXT) of Japan.
The authors thank the Yukawa Institute for Theoretical Physics at Kyoto University. 
Discussions during the GCOE/YITP workshop YITP-W-09-01  on "Non-linear cosmological perturbations" 
were useful to complete this work. 
\end{acknowledgments}

\appendix
\section{Correspondence in the continuum limit} \label{D and C}
In this paper, to tame the divergence in the IR limit, 
we begin with the model with a finite volume of
the universe $V_c\equiv L_c^3$. 
After we define appropriate observables, we take the
infinite volume limit $V_c \rightarrow \infty$. 
At this step the discrete label of the co-moving wave number changes 
to the continuum one. 
We use two different notations for the Fourier components 
between the perturbation variable like $\varphi^I(x)$ and the window function
$\WL(\bm{x})$. 
Because of that, these two classes of quantities should be treated 
differently when we take the continuum limit. 
Hereafter, we use variables $Q$ and $W$ to represent the variables 
of the first and the second classes, respectively. 
We add the index $(d)$ on the discrete Fourier components in the finite volume, 
while the index $(c)$ on the continuous ones in the infinite volume.
If not necessary, we abbreviate the suffixes, $I$ and $\alpha$. 

\subsection{Quantized variables}

When we consider the first class of variables $Q(x)$, which is 
to be quantized like $\varphi(x)$, 
the corresponding mode functions $\{q_{\sbm{k}}\}$ play the more
important role than the Fourier components of $Q(x)$ themselves. 
Therefore we adopt a convention so that
$q_{\sbm{k}}$ remains unchanged in the continuum limit  $V_c ~\rightarrow ~\infty$.

When we quantize $Q(x)$, 
we expand the Fourier mode $Q_k$ in terms of the creation and
annihilation operator like 
\begin{eqnarray}
 Q_{\sbm{k}} = (2 \pi)^{\frac{3}{2}} \{a_{\sbm{k}}q_{\sbm{k}} +
  a^{\dagger}_{\sbm{k}}q^*_{\sbm{k}} \}~.
\end{eqnarray}
We require that the mode functions $q_{\sbm{k}}$ remain 
unchanged in the continuum limit. 
On the other hand, 
the commutation relation for the creation and annihilation operator
\begin{eqnarray*}
 \left[a^{(d)}_{\sbm{k}}, a^{(d)}_{\sbm{k}'}{}^{\dagger}\right] 
   = \delta_{\sbm{k},\sbm{k}'}
\end{eqnarray*}
changes in the continuum limit to  
\begin{eqnarray*}
 \left[ a^{(c)}_{\sbm{k}}, a^{(c)}_{\sbm{k}'}{}^{\dagger}\right] =
 \delta^3( \bm{k} - \bm{k}' )~. 
\end{eqnarray*}
From the above commutation relation, a trivial relations 
\begin{eqnarray*}
&& 1=\sum_{\sbm{k}'} 
    \left[a^{(d)}_{\sbm{k}},
		    a^{(d)}_{\sbm{k}'}{}^{\dagger}\right], 
\cr
&& 1=  \int d^3 k'\, 
    \left[a^{(c)}_{\sbm{k}},
		    a^{(c)}_{\sbm{k}'}{}^{\dagger}\right], 
\end{eqnarray*}
follow. 
Since the wave number $\bm{k}$ is discrete like   
$k_i = \Delta k j_i \equiv 2 \pi j_i/L_c$ with
$j_i$ being integer,  
$\sum_{\sbm{k}'}$ is to be replaced with 
$(2\pi)^{-3}V_c \int d^3k$
in the continuum limit. 
This requires the correspondence like  
\begin{eqnarray*} 
a^{(d)}_{\sbm{k}} \longleftrightarrow \bar{V}_c^{- \frac{1}{2}}~ a^{(c)}_{\sbm{k}}, 
\end{eqnarray*}
where $\bar{V}_c \equiv V_c / (2 \pi)^3$ and hence we find 
\begin{equation}
 Q_{\sbm{k}}^{(d)} 
\longleftrightarrow \bar{V}_c^{- \frac{1}{2}} Q_{\sbm{k}}^{(c)}. 
\end{equation}

To realize the above correspondence, 
we define the Fourier components $Q_{\sbm{k}}^{(d)}$ as 
\begin{eqnarray}
 Q_{\sbm{k}}^{(d)} =  \frac{1}{\bar{V}_c^{\frac{1}{2}}} \int_{V_c} d^3 x\,
  Q(x) e^{- i \sbm{k} \cdot \sbm{x}} ~.  \label{mode exp qkuc}
\end{eqnarray}
Then the definition of the Fourier components in the continuum 
limit becomes a normal one:
\begin{eqnarray}
 Q_{\sbm{k}}^{(c)} =  \int d^3 x\,
  Q(x) e^{- i \sbm{k} \cdot \sbm{x}} ~.  
\end{eqnarray}
The inverse transform is given by 
\begin{eqnarray}
 Q(x) = \frac{1}{V_c^{\frac{1}{2}} (2 \pi)^{\frac{3}{2}}} \sum_{\sbm{k}} Q_{\sbm{k}}^{(d)}
 e^{i \sbm{k} \cdot \sbm{x}},  
\end{eqnarray}
and its continuum limit consistently recovers 
\begin{eqnarray}
 Q(x) & = & \frac{1}{(2 \pi)^3} \sum_{\sbm{k}} (\Delta k)^3
 \bar{V}_c^{\frac{1}{2}}~ Q_{\sbm{k}}^{(d)} e^{i \sbm{k} \cdot \sbm{x}} 
\label{A6'}
\\
 &\rightarrow &
 \frac{1}{(2 \pi)^3} \int d^3k\, Q_{\sbm{k}}^{(c)} 
 e^{i \sbm{k} \cdot \sbm{x}}.
\label{A6}
\end{eqnarray}

In \S \ref{Bogoliubov}, we treat the $\bm{k}=0$ mode separately. 
Even in the limit $V_c\to\infty$, this mode is treated as a 
discrete spectrum. Hence, the corresponding commutators 
keep the normalization in terms of Kronecker $\delta$, 
and we have 
\begin{eqnarray*}
 \left[a^{(c)}_{~\baralpha,0}, a^{(c)}_{~\baralpha,0}{}^{\dagger}\right] 
   = 1.
\end{eqnarray*}
This means that the correspondence of the 
creation and annihilation operators should be like
$a^{(c)}_{~\baralpha,0} \leftrightarrow a^{(d)}_{~\baralpha,0}$. 
Taking into account the exception for the $\bm{k}=0$ mode, 
Eq.~(\ref{A6}) is to be modified to 
\begin{eqnarray}
 Q_{\baralpha}(x) = {Q_{~\baralpha,0}^{(c)} \over (2 \pi)^{\frac{3}{2}} }
 + \frac{1}{(2 \pi)^3} \int d^3k\, Q_{\baralpha, \sbm{k}}^{(c)} 
 e^{i \sbm{k} \cdot \sbm{x}}.
\end{eqnarray}
where we have denoted the contribution from the basis $\baralpha$.
Comparing this with the discrete case (\ref{A6'}), we find that we
should have 
\begin{eqnarray*}
 Q_{~\baralpha,0}^{(c)}
 \longleftrightarrow V_c^{- \frac{1}{2}} Q_{~\baralpha,0}^{(d)}.
\end{eqnarray*}
Correspondingly, we have 
\begin{eqnarray*}
 q_{~\baralpha,0}^{(c)}
 \longleftrightarrow V_c^{- \frac{1}{2}} q_{~\baralpha,0}^{(d)}.
\end{eqnarray*}

Finally, we mention the normalization conditions of mode functions. 
The normalization conditions for $q_{\sbm{k}}^{(d)}$ 
given in Eq.~(\ref{normalization}) are understood as 
\begin{eqnarray}
 \left(q_{\sbm{k}}^{(d)}, q_{\sbm{k}'}^{(d)}\right) & = & V_c\,
  \delta_{\sbm{k},\sbm{k}'} 
=\int d^3 x\, e^{i(\sbm{k}-\sbm{k}')\cdot \sbm{x}}. 
\end{eqnarray}
Hence, in the limit $V_c\to\infty$ the right hand side 
is to be understood as $(2\pi)^3\delta^3(\bm{k}-\bm{k}')$. 
Therefore the normalization conditions of mode functions 
in the continuum limit should be 
\begin{eqnarray}
 \left(q_{\sbm{k}}^{(c)}, q_{\sbm{k}'}^{(c)}\right) & = & (2\pi)^3
  \delta^3 (\bm{k}-\bm{k}')~.  
\end{eqnarray}

To summarize, the correspondence between the discrete Fourier components 
and the Fourier components in the continuum limit is given by 
\begin{eqnarray}
 Q^{(c)}_{\sbm{k}} \leftrightarrow \bar{V}_c^{\frac{1}{2}} Q^{(d)}_{\sbm{k}}, ~~~
 q^{(c)}_{\sbm{k}} \leftrightarrow  q^{(d)}_{\sbm{k}}, ~~~
 a^{(c)}_{\sbm{k}} \leftrightarrow  \bar{V}_c^{\frac{1}{2}} a^{(d)}_{\sbm{k}}, 
\end{eqnarray}
for all modes $\bm{k}$ with $\alpha \ne \baralpha$ and $\bm{k} \ne 0$
with $\alpha = \baralpha$ and
\begin{eqnarray}
 Q^{(c)}_{~\baralpha, 0} \leftrightarrow  V_c^{- \frac{1}{2}}
  Q^{(d)}_{~\baralpha, 0}, ~~~
 q^{(c)}_{~ \baralpha, 0} \leftrightarrow  V_c^{- \frac{1}{2}} q^{(d)}_{~ \baralpha,0}, ~~~
 a^{(c)}_{~ \baralpha,0} \leftrightarrow  a^{(d)}_{~
 \baralpha,0}. \nonumber \\
\end{eqnarray}

\subsection{Un-quantized variables}
The quantities of the second class, $W(x)$, 
are not supposed to be quantized. 
In this case it is more convenient to consider  
the Fourier components $W_{\sbm{k}}$ that 
remain unchanged in the continuum limit:
\begin{eqnarray*}
  W_{\sbm{k}}^{(c)} \longleftrightarrow  W_{\sbm{k}}^{(d)}. 
\end{eqnarray*} 

For this purpose, we simply define the Fourier components in 
a usual manner by 
\begin{eqnarray}
 W_{\sbm{k}}^{(d)} 
  =   \int_{V_c} d^3 x\,
  W(x) e^{- i \sbm{k} \cdot \sbm{x}}.  \label{mode exp qkc}
\end{eqnarray}
Then the inverse transform is given by 
\begin{eqnarray}
  W(x) = \frac{1}{V_c} \sum_{\sbm{k}} W_{\sbm{k}}^{(d)}
 e^{i \sbm{k} \cdot \sbm{x}},  
\end{eqnarray}
and its continuum limit $V_c \rightarrow \infty$ becomes 
\begin{eqnarray}
 W(x) &=& \frac{1}{(2 \pi)^3} \sum_{\sbm{k}} (\Delta k)^3
   W_{\sbm{k}}^{(d)} e^{i \sbm{k} \cdot \sbm{x}} 
\cr &
\rightarrow &
  \frac{1}{(2 \pi)^3} \int d^3k\, W_{\sbm{k}}^{(c)} 
 e^{i \sbm{k} \cdot \sbm{x}},   
\end{eqnarray}
as is expected.

\section{Bogoliubov transformation} \label{Sec:Bogoliubov}
\subsection{Another set of mode functions} \label{Bogoliubov}
In Sec. \ref{Squeezed}, we transformed the mode functions to more suitable
ones to understand the role of the projection operator. 
In this subsection, we give the detailed explanations for the two transformations.
On the first transformation, the new basis mode functions $\{v^I_{~ \bar{\alpha}, {\sbm p}}\}$ are constructed 
so that the leading amplitude in the long wave-length limit is canceled
for $\bm{p} \neq 0$. While, the IR divergent contributions are
localized to a single mode with $p=0$.
Hence, when we expand $\tilde\varphi^I_n$ 
in terms of the
creation and annihilation operators associated with 
the new set of mode functions, the expansion coefficients 
$C[\tilde\varphi^I_n]$ are IRVSFs except for the case with $p_j=0$. 
Such transformation can be achieved by taking
\begin{eqnarray}
 v^I_{\,\baralpha,0} (x) &\equiv&
 \frac{1}{{\cal N}_{\baralpha}} \biggl\{ u^I_{\,\baralpha,0} (t)  \nonumber \\ && \quad
 \quad \quad
 + \sum_{\sbm{p} \neq 0}
 \frac{W_{-\sbm{p}}{}^*}{W_0^* }
 \frac{c^*_{\baralpha}(0)}{c^*_{\baralpha}(p)}
 u^I_{\,\baralpha, p} (t) e^{i \sbm{p} \cdot \sbm{x}} \biggr\}~,
 \nonumber \\ \label{def v0} \\ 
 v^I_{\,\baralpha,\sbm{p}} (x) &\equiv& u^I_{\, \baralpha, p}(t) e^{i \sbm{p} \cdot \sbm{x}} 
    - \frac{W_{- \sbm{p}} }{W_0}
 \frac{c_{\baralpha}(0)}{c_{\baralpha}(p)}\,u^I_{\,\bar\alpha,0} (t).  \label{def vk}
\end{eqnarray}
where we denote $\WL_{\sbm p}$ by $W_{\sbm p}$.
The normalization constant ${\cal N}_{\baralpha}$ is chosen as 
\begin{eqnarray}
&&{\cal N}_{\baralpha}^2 = \sum_{\sbm{p}}
 \left\vert\frac{W_{- \sbm{p}}}{W_0} \frac{c_{\baralpha}(0)}{c_{\baralpha}(p)}
  \right\vert^2 
  = 1 + O(s_{\baralpha})~,
\end{eqnarray}
where the second equality immediately follows from the observation 
that only the term with $\bm{p}=0$ remains in
the limit $s_{\baralpha} \rightarrow 0$.

The orthonormal relation for $\{ v^I_{~ \baralpha, \sbm{k}} \}$ is given by
\begin{eqnarray}
 (v_{~\baralpha,0},  v_{\beta,0}) &=& V_c \delta_{\alpha\beta},
        \nonumber    \\
 (v_{~\baralpha,0},  v_{\beta, \sbm{p}}) & =&  0,    \nonumber   \\
 (v_{~\baralpha, \sbm{p}}, v_{\beta, \sbm{p}'} )
 & = & V_c \delta_{\alpha\beta}\left\{ \delta_{\sbm{p}, \sbm{p}'}
  + \frac{W_{- \sbm{p}} W^*_{-\sbm{p}'} 
    |c_{\baralpha}(0)|^2}
    {|W_0|^2 {c_{\baralpha}(p)} c^*_{\baralpha}(p')} 
\right\} 
\cr & = & V_c \delta_{\alpha\beta}
 \Bigl\{ \delta_{\sbm{p}, \sbm{p}'}
  + O( s_{\baralpha}^2 / V_c) \Bigr\},             \nonumber   \\
 (v_{~\baralpha,0},  v_{\beta,0}^*) 
 & = &
  (v_{~\baralpha,0},  v_{\beta, \sbm{p}}^*)
 = (v_{~ \baralpha, \sbm{p}}, v_{\beta, \sbm{p}'}^* ) =
 0~. \label{ON vkvk}
\end{eqnarray}
The mode functions $v^I_{~ \baralpha, \sbm{p} \neq 0}$ are not 
mutually orthogonal before we take the limit $s_{\baralpha} \to 0$ in Eq.~(\ref{ON vkvk}). 
After taking the limit $s_{\baralpha} \to 0$, however, 
$\{v^I_{~ \baralpha, \sbm{p}}\}$ becomes a set of orthonormal bases. 
We denote the creation
and annihilation operators associated with $\{v^I_{~ \baralpha, \sbm{p}}\}$ by 
$b_{\baralpha, \sbm{p}}^\dagger$ and $b_{\baralpha, \sbm{p}}$,
respectively. 

After this transformation, the IR divergent contribution  is confined to 
$v^I_{~\baralpha,0}(x)$. Indeed, 
the problematic term in $u^I_{~\baralpha,p}$ which scales as
$p^{-(3 + \delta_{\baralpha})/2}$ in the IR limit is cancelled by the second term 
in Eq.~(\ref{def vk}). Then, even when $\delta_{\baralpha} > 0 $,  
as long as $\delta_\alpha <1$, 
we find that $p^{\frac{3}{2}} v^I_{~\baralpha, {\sbm p} \neq 0}$
vanishes in the limit ${\bm p} \rightarrow 0$. This is the necessary
and sufficient condition for the coefficients of the creation and annihilation
operators to be IRVSFs. In contrast, $v^I_{~\baralpha,0}$ has a 
diverging amplitude in the limit $s_{\bar \alpha} \rightarrow 0$ 
and $V_c \rightarrow  \infty$.
Notice that this transformation does not mix the positive
frequency modes with the negative frequency ones. 
Therefore in the limit $s_{\baralpha} \to 0$ the vacuum annihilated
by $b_{\baralpha, \sbm{p}}$ is identical 
to the vacuum annihilated by $a^{\baralpha}_{\sbm{p}}$.

\begin{figure}[t]
\begin{center}
\includegraphics[width=8cm]{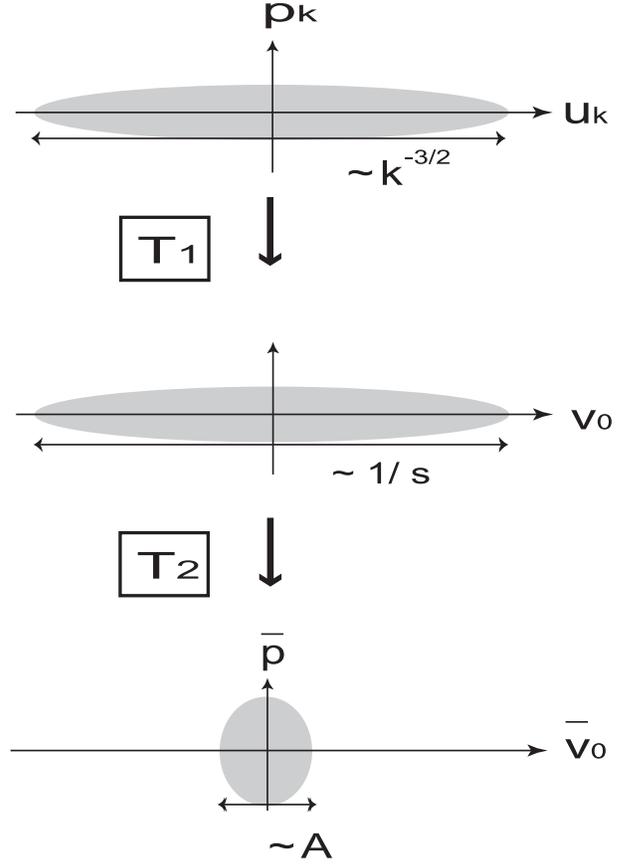}
\caption{Two Bogoliubov transformations  \label{fg:Bogoliubov}}
\end{center}
\end{figure} 
%
Now we move on to the second step of transformation. 
Here we introduce another mode function 
for $p=0$ mode which 
naturally defines wave packets with a finite width
even in the limit $s_{\bar \alpha} \rightarrow 0$ and $V_c \rightarrow
\infty$. 
In Fig.~\ref{fg:Bogoliubov}, we depicted the changes of the wave
packets under these two Bogoliubov transformations. For brevity, we have
suppressed the indices $I$ and $\alpha$. 
We denote a new set of mode functions by
$\{ \bar{v}^I_{~ \baralpha, \sbm{p}} \}$,  
which is defined by 
\begin{eqnarray}
 \bar{v}^I_{~\baralpha,0} (x) &\!=\!& \cosh r_{\baralpha}~  v^I_{~\baralpha,0}(x) -
  \sinh r_{\baralpha}~ v^{I~~*}_{~\baralpha,0} (x), \label{def barv0} \\
 \bar{v}^I_{~\baralpha, \sbm{k}} (x) &\!=\!& v^I_{~\baralpha, \sbm{k}} (x), \label{def barvk}
\end{eqnarray}
with the squeeze parameter $r_{\baralpha}$ such that
\begin{eqnarray}
 e^{r_{\baralpha}} = \frac{1}{{\cal N}_{\baralpha}\!A_{\baralpha}s_{\baralpha}} ~,  \label{squeeze parameterap}
\end{eqnarray}
where $A_{\baralpha}$ is a real constant. 
The mode functions for ${\bm p}\ne 0$ 
are unchanged:
\begin{eqnarray}
\bar{v}^I_{~\baralpha, \sbm{p}} = v^I_{~ \baralpha, \sbm{p}}, 
\quad \mbox{for}~{\bm p}\ne 0.
\end{eqnarray}
We denote the creation and annihilation operators associated with $\{\bar{v}^I_{~ \baralpha, \sbm{p}}\}$ by 
$\bara_{\baralpha\sbm{p}}^\dagger$ and $\bara_{\baralpha \sbm{p}}$.

In the limit $s_{\baralpha} \rightarrow 0$, we obtain 
\begin{eqnarray}
 \bar{v}^I_{\,\baralpha,0} (x) &=& V_c^{\frac{1}{2}} \Biggl[
 A_{\baralpha} g^I_{\,\baralpha,0}(t) + \frac{i}{ A_{\baralpha}} \frac{1}{V_c}
 \,d^I_{\,\baralpha,0}(t)
 \cr && \qquad \qquad 
 + \frac{i}{A_{\baralpha}}  \frac{1}{V_c}  \sum_{\sbm{p} \neq 0}
  \frac{W_{\sbm p}}{W_0} 
  d^I_{\,\baralpha,p}(t) e^{i \sbm{p} \cdot \sbm{x}} \Biggr]~,
 \nonumber \\  \label{v0b}
\end{eqnarray}
The first and the second terms in Eq.~(\ref{v0b}) 
originate from the homogeneous mode, $u^I_{~\baralpha,0}$, and its complex conjugate. 
The terms that diverge in the limit $s_{\baralpha}\to 0$ are common 
in $u^I_{~\baralpha,0}$ and $u^I_{~\baralpha,0}{}^*$, which 
cancel with each other in the 
definition of $\bar v^I_{~\baralpha,0}$ in Eq.~(\ref{def barv0}). 
The third term in Eq.~(\ref{v0b}) originates from the inhomogeneous modes,
$u^I_{\,\alpha, p \ne 0}$. 
The contribution from $u^I_{~\alpha, p \ne 0}$ is also large in 
$v^I_{\,\alpha,0}$ 
in the IR limit, but it is not in $\bar v^I_{\,\alpha,0}$ .
The leading terms of $d^I_{~\baralpha,p}(t)$ in ${\bm p} \rightarrow 0$
limit behaves as 
\begin{eqnarray*}
d^I_{~\baralpha,p}(t) \approx d^I_{~\baralpha}(t)~. 
\end{eqnarray*}
Hence, 
the momentum integral $\int d^3 \bm{p}\, 
W_{\sbm p} d^I_{\,\baralpha,p}(t) e^{i{\sbm p}\cdot{\sbm x}}$, 
which arises in the continuum limit, 
remains finite for $x<\infty$.

Here, we choose the unspecified constant 
$A_{\baralpha}$ such that 
the mode functions $\bar v^I_{\baralpha,0}$
take the minimum amplitude on average. 
To do so, we compare the amplitude of the first term with that of 
the last term in Eq.~(\ref{v0b}). 
Near the final time $t_f$ the last term is negligibly  
small, but it is larger in the past.  
Assuming $\dot\phi$ and $H$ are almost constant, 
the relative amplitude between these two terms 
is roughly estimated as 
$|A_\baralpha^{-2}\int_0^{aH} d^3p\, d^I_{~\baralpha,p}|
\approx H^2/A_\baralpha^2$, 
where we used  $
d^I_{~\baralpha,p}\approx 1/(6Ha^3)\delta^I_{\,\baralpha}$. 
Hence, we find that it is appropriate to choose 
$A_\balpha = H$ 
to minimize the amplitude of the mode functions. 
(We choose $A_1 = 1/s_1 = \infty$ so that the mode function for the 
adiabatic mode is unchanged:
$\bar v^I_{1,0}=v^I_{1,0}$.)

Now, sending $V_c \rightarrow \infty$, we take the continuum limit. 
We have summarized the correspondence between the discrete description and
the continuous one in Appendix \ref{D and C}. 
The overall factor
$V_c^{\frac{1}{2}}$ in Eq.~(\ref{v0b}) is cancelled by $V_c^{-
\frac{1}{2}}$ in Eq.(\ref{Expansion:varphiI}). 
In this limit, $\sum_k$ is replaced with 
$(2\pi)^{-3} V_c \int d^3k$.  
Since the expression (\ref{v0b}) has a factor $1/V_c$ 
in front of the summation $\sum_k$, we find that 
$\bar v^I_{~\baralpha,0}$ is free from the explicit 
divergence in the limit $V_c\to \infty$. In the continuum limit, 
$\breve{\varphi}^I_1$ can be expanded as
\begin{eqnarray}
\breve{\varphi}^I_1(x) \!&=\!& \sum_{\baralpha} \left\{ {\bar{v}^I_{~\baralpha,0} \over \Mp}
  \bara_{\baralpha,0} + 
 \int_{\sbm{p} \ne 0}\! \frac{d^3 {\bm p}}{(2 \pi)^{\frac{3}{2}}} 
  {\bar{v}^I_{~\baralpha, \sbm{p}}(t)\over M_{\rm pl}}
  \bara_{\alpha,{\sbm p}}  \right\}\nonumber\\
 &&\qquad\qquad + (\mbox{h.c.})~,
\label{Def:varphisap}
\end{eqnarray}
where
\begin{eqnarray}
 \bar{v}^I_{\,\baralpha,0} (x) \!&=\!& 
 H_f g^I_{\,\baralpha,0}(t) 
 + \frac{i}{H_f} \int_{\sbm{p} \ne 0} \frac{d^3 \bm{p}}{(2 \pi)^3}
  \frac{W_{\sbm p}}{W_0} 
  d^I_{\,\baralpha,p}(t) e^{i \sbm{p} \cdot \sbm{x}}~, \nonumber \\ 
 \label{v0kbap}\\
 \bar{v}^I_{\, \baralpha,\sbm{p}} (x) \!&=\!& u^I_{\, \baralpha, p}(t) e^{i \sbm{p} \cdot \sbm{x}} 
    - \frac{W_{- \sbm{p}} }{W_0} {g^I_{\,\balpha,0} (t) \over
    c_{\baralpha}(p)}~. 
 \label{vpkbap}
\end{eqnarray}

\subsection{Expansion by coherent states} 
In this subsection, we show that the adiabatic vacuum state 
$|\,0\,\rangle_a$ can be expanded in terms of the coherent 
states $|\beta\rangle_{\bara}$ associated with $\bara_{\baralpha,0}$. 
The coherent state $|\beta\rangle_{\bara}$ is 
defined by 
\begin{eqnarray}
 |\beta \rangle_{\bara} \equiv \prod_{\balpha=2}^D
  e^{\beta_{\balpha} ( \bara_{\balpha, 0}^{\dagger} - 
  \bara_{\balpha, 0} )}
|\,0\, \rangle_{\bara}
 = \prod_{\balpha=2}^D e^{- \frac{ \beta_{\balpha}^2}{2}}
 e^{\beta_{\balpha} \bara_{\balpha,0}^\dagger}
 \,|\, 0\, \rangle_{\bara}~, \nonumber \\
\label{def coherentap}
\end{eqnarray}
where $|\,0\, \rangle_{\bara}$ 
is the vacuum state annihilated by 
$\bara_{\balpha,\sbm{k}}$. This new vacuum state 
$|\,0\, \rangle_{\bara}$ 
is related to $|\,0\, \rangle_{a}$ by 
\begin{eqnarray}
 |\,0\, \rangle_a = \prod_{\balpha=2}^D (\cosh r_{\balpha})^{-
  1/2} e^{ \frac{1}{2}\! \tanh r_{\balpha}\,
  (\bara_{\balpha,0}^\dagger)^2} 
 \,|\, 0\, \rangle_{\bara} ~. \label{relation 0a 0bb2ap}
\end{eqnarray}
It is easy to verify that the left hand side is annihilated by 
$b_{\balpha,0}=(\cosh r_{\balpha})\, \bara_{\balpha,0} 
 - (\sinh r_{\balpha})\, \bara_{\balpha,0}^\dagger$. 
The coherent state satisfies
\begin{eqnarray*} 
 && \bara_{\balpha,\sbm {p}} |\beta \rangle_{\bara} =
  \beta_{\balpha} |\beta \rangle_{\bara}~, \cr
  && \bara_{1,\sbm {p}} |\beta \rangle_{\bara}  =0~.
\end{eqnarray*}
The original vacuum state $|\,0\, \rangle_a $  
can be expressed as
a superposition of the new coherent states like 
\begin{eqnarray}
 |\,0\, \rangle_a 
 &=& \prod_{\balpha=2}^D \left[ \int^{\infty}_{-\infty} 
     d\beta_{\balpha}\, E_{\balpha}(\beta_{\balpha}) \right] \,|\beta \rangle_{\bara}~, 
  \label{BD by csap}
\end{eqnarray}
where the coefficient $E_{\balpha}({\beta_{\balpha}})$ is given by
\begin{eqnarray}
 E_{\balpha}(\beta_{\balpha}) &\equiv&  (2 \pi  \sinh r_{\balpha})^{- 1/2}
 e^{- \beta_{\balpha}^2  (e^{2r_{\balpha}} - 1 )^{-1}}\cr
&\to &
\sqrt{s_{\balpha} H_f \over \pi}e^{-(s_{\balpha} H_f \beta_{\balpha})^2},~
 (s_{\balpha} \to 0).~  
\label{csdefap}
\end{eqnarray}
The formula (\ref{BD by csap}) can be verified simply by performing
Gaussian integral about $\{ \beta_{\balpha} \}$. %



\begin{thebibliography}{99}
\bibitem{Boyanovsky:2004gq}
  D.~Boyanovsky and H.~J.~de Vega,
  Phys.\ Rev.\ D {\bf 70}, 063508 (2004)
  [arXiv:astro-ph/0406287].

\bibitem{Boyanovsky:2004ph}
  D.~Boyanovsky, H.~J.~de Vega and N.~G.~Sanchez,
  Phys.\ Rev.\ D {\bf 71}, 023509 (2005)
  [arXiv:astro-ph/0409406].

\bibitem{Boyanovsky:2005sh}
  D.~Boyanovsky, H.~J.~de Vega and N.~G.~Sanchez,
  Nucl.\ Phys.\ B {\bf 747}, 25 (2006)
  [arXiv:astro-ph/0503669].

\bibitem{Boyanovsky:2005px}
  D.~Boyanovsky, H.~J.~de Vega and N.~G.~Sanchez,
  Phys.\ Rev.\ D {\bf 72}, 103006 (2005)
  [arXiv:astro-ph/0507596].

\bibitem{Onemli:2002hr}
  V.~K.~Onemli and R.~P.~Woodard,
  Class.\ Quant.\ Grav.\  {\bf 19}, 4607 (2002)
  [arXiv:gr-qc/0204065].

\bibitem{Brunier:2004sb}
  T.~Brunier, V.~K.~Onemli and R.~P.~Woodard,
  Class.\ Quant.\ Grav.\  {\bf 22}, 59 (2005)
  [arXiv:gr-qc/0408080].

\bibitem{Prokopec:2007ak}
  T.~Prokopec, N.~C.~Tsamis and R.~P.~Woodard,
  arXiv:0707.0847 [gr-qc].

\bibitem{Sloth:2006az}
  M.~S.~Sloth,
  Nucl.\ Phys.\ B {\bf 748}, 149 (2006)
  [arXiv:astro-ph/0604488].

\bibitem{Sloth:2006nu}
  M.~S.~Sloth,
  Nucl.\ Phys.\ B {\bf 775}, 78 (2007)
  [arXiv:hep-th/0612138].

\bibitem{Seery:2007we}
  D.~Seery,
  JCAP {\bf 0711}, 025 (2007)
  [arXiv:0707.3377 [astro-ph]].

\bibitem{Seery:2007wf}
  D.~Seery,
  JCAP {\bf 0802}, 006 (2008)
  [arXiv:0707.3378 [astro-ph]].

\bibitem{Urakawa:2008rb}
  Y.~Urakawa and K.~i.~Maeda,
  Phys.\ Rev.\  D {\bf 78}, 064004 (2008)
  [arXiv:0801.0126 [hep-th]].



\bibitem{Komatsu:2008hk}
  E.~Komatsu {\it et al.}  [WMAP Collaboration],
  arXiv:0803.0547 [astro-ph].



\bibitem{Bartolo:2001cw}
  N.~Bartolo, S.~Matarrese and A.~Riotto,
  Phys.\ Rev.\  D {\bf 65}, 103505 (2002)
  [arXiv:hep-ph/0112261].

\bibitem{Bartolo:2004if}
  N.~Bartolo, E.~Komatsu, S.~Matarrese and A.~Riotto,
  Phys.\ Rept.\  {\bf 402}, 103 (2004)
  [arXiv:astro-ph/0406398].

\bibitem{Maldacena:2002vr}
  J.~M.~Maldacena,
  JHEP {\bf 0305}, 013 (2003)
  [arXiv:astro-ph/0210603].

\bibitem{Kim:2006te}
  S.~A.~Kim and A.~R.~Liddle,
  Phys.\ Rev.\  D {\bf 74}, 063522 (2006)
  [arXiv:astro-ph/0608186].

\bibitem{Babich:2004gb}
  D.~Babich, P.~Creminelli and M.~Zaldarriaga,
  JCAP {\bf 0408}, 009 (2004)
  [arXiv:astro-ph/0405356].

\bibitem{Seery:2005wm}
  D.~Seery and J.~E.~Lidsey,
  JCAP {\bf 0506}, 003 (2005)
  [arXiv:astro-ph/0503692].

\bibitem{Seery:2005gb}
  D.~Seery and J.~E.~Lidsey,
  JCAP {\bf 0509}, 011 (2005)
  [arXiv:astro-ph/0506056].

\bibitem{Weinberg:2005vy}
  S.~Weinberg,
  Phys.\ Rev.\ D {\bf 72}, 043514 (2005)
  [arXiv:hep-th/0506236].

\bibitem{Weinberg:2006ac}
  S.~Weinberg,
  Phys.\ Rev.\ D {\bf 74}, 023508 (2006)
  [arXiv:hep-th/0605244].
  
\bibitem{Rigopoulos:2005xx}
  G.~I.~Rigopoulos, E.~P.~S.~Shellard and B.~J.~W.~van Tent,
  Phys.\ Rev.\  D {\bf 73}, 083521 (2006)
  [arXiv:astro-ph/0504508].

\bibitem{Rigopoulos:2005ae}
  G.~I.~Rigopoulos, E.~P.~S.~Shellard and B.~J.~W.~van Tent,
  Phys.\ Rev.\  D {\bf 73}, 083522 (2006)
  [arXiv:astro-ph/0506704].


\bibitem{Rigopoulos:2005us}
  G.~I.~Rigopoulos, E.~P.~S.~Shellard and B.~J.~W.~van Tent,
  Phys.\ Rev.\  D {\bf 76}, 083512 (2007)
  [arXiv:astro-ph/0511041].

\bibitem{Vernizzi:2006ve}
  F.~Vernizzi and D.~Wands,
  JCAP {\bf 0605}, 019 (2006)
  [arXiv:astro-ph/0603799].

\bibitem{Chen:2006nt}
  X.~Chen, M.~x.~Huang, S.~Kachru and G.~Shiu,
  JCAP {\bf 0701}, 002 (2007)
  [arXiv:hep-th/0605045].

\bibitem{Battefeld:2006sz}
  T.~Battefeld and R.~Easther,
  JCAP {\bf 0703}, 020 (2007)
  [arXiv:astro-ph/0610296].

\bibitem{Yokoyama:2007dw}
  S.~Yokoyama, T.~Suyama and T.~Tanaka,
  Phys.\ Rev.\  D {\bf 77}, 083511 (2008)
  [arXiv:0711.2920 [astro-ph]].

\bibitem{Yokoyama:2008by}
  S.~Yokoyama, T.~Suyama and T.~Tanaka,
  JCAP {\bf 0902}, 012 (2009)
  [arXiv:0810.3053 [astro-ph]].


\bibitem{Seery:2008ax}
  D.~Seery, M.~S.~Sloth and F.~Vernizzi,
  arXiv:0811.3934 [astro-ph].

\bibitem{Naruko:2008sq}
  A.~Naruko and M.~Sasaki,
  Prog.\ Theor.\ Phys.\  {\bf 121}, 193 (2009)
  [arXiv:0807.0180 [astro-ph]].

\bibitem{Weinberg:2008mc}
  S.~Weinberg,
  arXiv:0805.3781 [hep-th].

\bibitem{Weinberg:2008nf}
  S.~Weinberg,
  Phys.\ Rev.\  D {\bf 78}, 123521 (2008)
  [arXiv:0808.2909 [hep-th]].

\bibitem{Weinberg:2008si}
  S.~Weinberg,
  arXiv:0810.2831 [hep-ph].

\bibitem{Cogollo:2008bi}
  H.~R.~S.~Cogollo, Y.~Rodriguez and C.~A.~Valenzuela-Toledo,
  JCAP {\bf 0808}, 029 (2008)
  [arXiv:0806.1546 [astro-ph]].


\bibitem{Rodriguez:2008hy}
  Y.~Rodriguez and C.~A.~Valenzuela-Toledo,
  arXiv:0811.4092 [astro-ph].


\bibitem{Lyth:2007jh}
  D.~H.~Lyth,
  JCAP {\bf 0712}, 016 (2007)
  [arXiv:0707.0361 [astro-ph]].

\bibitem{Bartolo:2007ti}
  N.~Bartolo, S.~Matarrese, M.~Pietroni, A.~Riotto and D.~Seery,
  JCAP {\bf 0801}, 015 (2008)
  [arXiv:0711.4263 [astro-ph]].

\bibitem{Riotto:2008mv}
  A.~Riotto and M.~S.~Sloth,
  JCAP {\bf 0804}, 030 (2008)
  [arXiv:0801.1845 [hep-ph]].

\bibitem{Enqvist:2008kt}
  K.~Enqvist, S.~Nurmi, D.~Podolsky and G.~I.~Rigopoulos,
  JCAP {\bf 0804}, 025 (2008)
  [arXiv:0802.0395 [astro-ph]].

\bibitem{Urakawa:2009my}
  Y.~Urakawa and T.~Tanaka,
  arXiv:0902.3209 [hep-th].

\bibitem{Polarski:1995jg}
  D.~Polarski and A.~A.~Starobinsky,
  Class.\ Quant.\ Grav.\  {\bf 13}, 377 (1996)
  [arXiv:gr-qc/9504030].

\bibitem{Kiefer:2006je}
  C.~Kiefer, I.~Lohmar, D.~Polarski and A.~A.~Starobinsky,
  Class.\ Quant.\ Grav.\  {\bf 24}, 1699 (2007)
  [arXiv:astro-ph/0610700].

\bibitem{Starobinsky:1986fx}
  A.~ A.~ Starobinsky, Lect. Notes Phys. {\bf 246}, 107 (1986).

\bibitem{Starobinsky:1994bd}
  A.~A.~Starobinsky and J.~Yokoyama,
  Phys.\ Rev.\  D {\bf 50}, 6357 (1994)
  [arXiv:astro-ph/9407016].


\bibitem{Nakao:1988yi}
  K.~i.~Nakao, Y.~Nambu and M.~Sasaki,
  Prog.\ Theor.\ Phys.\  {\bf 80}, 1041 (1988).

\bibitem{Nambu:1988je}
  Y.~Nambu and M.~Sasaki,
  Phys.\ Lett.\  B {\bf 219}, 240 (1989).



\bibitem{Morikawa:1989xz}
  M.~Morikawa,
  Phys.\ Rev.\  D {\bf 42}, 1027 (1990).

\bibitem{Morikawa:1987ci}
  M.~Morikawa,
  Prog.\ Theor.\ Phys.\  {\bf 77}, 1163 (1987).



\bibitem{Tanaka:1997iy}
  T.~Tanaka and M.~a.~Sakagami,
  Prog.\ Theor.\ Phys.\  {\bf 100}, 547 (1998)
  [arXiv:gr-qc/9705054].

\bibitem{Seery:2009hs}
  D.~Seery,
  arXiv:0903.2788 [astro-ph.CO].

\bibitem{Gordon:2000hv}
  C.~Gordon, D.~Wands, B.~A.~Bassett and R.~Maartens,
  Phys.\ Rev.\  D {\bf 63}, 023506 (2001)
  [arXiv:astro-ph/0009131].

\bibitem{Lyth:2005du}
  D.~H.~Lyth and Y.~Rodriguez,
  Phys.\ Rev.\  D {\bf 71}, 123508 (2005)
  [arXiv:astro-ph/0502578].

\bibitem{Jarnhus:2007ia}
  P.~R.~Jarnhus and M.~S.~Sloth,
  JCAP {\bf 0802}, 013 (2008)
  [arXiv:0709.2708 [hep-th]].



\bibitem{Roura:2007jj}
  A.~Roura and E.~Verdaguer,
  arXiv:0709.1940 [gr-qc].

\bibitem{Urakawa:2007dm}
  Y.~Urakawa and K.~i.~Maeda,
  Phys.\ Rev.\  D {\bf 77}, 024013 (2008)
  [arXiv:0710.5342 [hep-th]].

\bibitem{Finelli:2008zg}
  F.~Finelli, G.~Marozzi, A.~A.~Starobinsky, G.~P.~Vacca and G.~Venturi,
  Phys.\ Rev.\  D {\bf 79}, 044007 (2009)
  [arXiv:0808.1786 [hep-th]].


%
%

\bibitem{Tsamis:1996qm}
  N.~C.~Tsamis and R.~P.~Woodard,
  Annals Phys.\  {\bf 253}, 1 (1997)
  [arXiv:hep-ph/9602316].

\bibitem{Tsamis:1996qq}
  N.~C.~Tsamis and R.~P.~Woodard,
  Nucl.\ Phys.\  B {\bf 474}, 235 (1996)
  [arXiv:hep-ph/9602315].

\bibitem{Garriga:2007zk}
  J.~Garriga and T.~Tanaka,
  Phys.\ Rev.\  D {\bf 77}, 024021 (2008)
  [arXiv:0706.0295 [hep-th]].

\bibitem{Tsamis:2007is}
  N.~C.~Tsamis and R.~P.~Woodard,
  arXiv:0708.2004 [hep-th].




\end{thebibliography}
\end{document}